\documentclass[prb,twocolumn]{revtex4-1}

\usepackage{mathtools}
\usepackage{amsmath,bm}
\usepackage{amssymb}
\usepackage{subfigure}
\usepackage{setspace}
\usepackage{dcolumn}
\usepackage{color}
\DeclareMathOperator{\sign}{sign}
\usepackage[hidelinks]{hyperref}

\begin{document}

\title{Edge States and Topological Pumping in Stiffness Modulated Elastic Plates}

\author{Emanuele Riva$^{a}$, Matheus I. N. Rosa$^{b}$ and  Massimo Ruzzene$^{c}$}
\affiliation{ $^a$ Department of Mechanical Engineering, Politecnico di Milano, Italy}
\affiliation{ $^b$ School of Mechanical Engineering, Georgia Institute of Technology, Atlanta GA 30332}
\affiliation{ $^c$ Department of Mechanical Engineering, University of Colorado Boulder, Boulder CO 80309}

\date{\today}

\begin{abstract}
We demonstrate that modulations of the stiffness properties of an elastic plate along a spatial dimension induce edge states spanning non-trivial gaps characterized by integer valued Chern numbers. We also show that topological pumping is induced by smooth variations of the phase of the modulation profile along one spatial dimension, which results in adiabatic edge-to-edge transitions of the edge states. The concept is first illustrated numerically for sinusoidal stiffness modulations, and then experimentally demonstrated in a plate with square-wave thickness profile. The presented numerical and experimental results show how continuous modulations of properties may be exploited in the quest for topological phases of matter. This opens new possibilities for topology-based waveguiding through slow modulations along a second dimension, spatial or temporal.
\end{abstract}

\maketitle

\section{Introduction}\label{Introduction}
The search for topological phases of matter has reached a mature state with multiple realizations across different physical realms, including quantum~\cite{hasan2010colloquium}, electromagnetic~\cite{lu2014topological,khanikaev2013photonic}, acoustic~\cite{PhysRevLett.114.114301,fleury2016floquet,lu2017observation} and elastic~\cite{mousavi2015topologically} media. In mechanics, topologically protected wave transport has been demonstrated through analogues to the \textit{Quantum Hall Effect} (QHE)~\cite{klitzing1980new,thouless1982quantized,prodan2009topological,wang2015topological,nash2015topological,souslov2017topological,mitchell2018amorphous,chen2019mechanical}, the \textit{Quantum Spin Hall Effect} (QSHE)~\cite{mousavi2015topologically,susstrunk2015observation,pal2016helical,PhysRevB.98.094302,chaunsali2018subwavelength,PhysRevX.8.031074} and the \textit{Quantum Valley Hall Effect} (QVHE)~\cite{pal2017edge,vila2017observation,liu2018tunable,liu2019experimental}. The rich underlying physics makes these robust waveguiding mechanisms promising for applications in acoustic devices or structural components designed to steer waves or isolate vibrations. 

Recently, topological phases have been explored in systems of lower physical dimensions by exploiting synthetic dimensions emerging from the exploration of relevant parameter spaces~\cite{qi2008topological,kraus2016quasiperiodicity,ozawa2016synthetic,lee2018electromagnetic}. Notable examples include the observation of edge states, commonly attributed to two-dimensional (2D) QHE systems, in one-dimensional (1D) quantum~\cite{alvarez2019edge}, electromagnetic~\cite{kraus2012topological}, acoustic~\cite{apigo2019observation,ni2019observation} and mechanical~\cite{apigo2018topological,rosa2019edge,Pal_2019} lattices following the Aubry-Andr\'{e}-Harper model of interactions~\cite{harper1955single,aubry1980analyticity}. Also, four-dimensional (4D) Quantum Hall phases have been realized using 2D photonic lattices~\cite{zilberberg2018photonic} and ultracold atoms~\cite{lohse2018exploring}, while six-dimensional (6D) phases in 3D systems have been theoretically investigated  in~\cite{petrides2018six,lee2018electromagnetic}. In this context, topological pumping has been pursued in a variety of physical systems, whereby adiabatic transitions of edge states are induced by smooth parameter variations along spatial~\cite{kraus2012topological,verbin2015topological,nakajima2016topological,lohse2016thouless,zilberberg2018photonic,lohse2018exploring,rosa2019edge} or temporal~\cite{grinberg2019robust,chen2019mechanical} dimensions. While previous experimental studies demonstrate pumping in photonic lattices and cold atomic gases, a realization using elastic waves is currently missing. 

In the quest for topological phases of matter, elastic solids such as thin, elastic plates are promising platforms due to the convenience they offer in terms of manufacturing and testing, and their rich spectral properties which are characterized by a large number of wave modes of distinct polarizations~\cite{PhysRevX.8.031074}. At the same time, the abundance of polarizations makes implementing topological waveguiding in elastic plates a challenging and non trivial development, when compared to acoustic~\cite{PhysRevLett.114.114301} and electromagnetic~\cite{lu2014topological,khanikaev2013photonic} counterparts. Towards overcoming these challenges and expanding the range of possibilities for topology-based elastic waveguiding, topological pumping is here experimentally demonstrated for the first time in a continuous elastic plate. The investigations herein leverage prior work on discrete lattices of continuous elastic waveguides~\cite{rosa2019edge} whereby modulations of physical properties along a spatial dimension were shown to induce edge states spanning non-trivial gaps. Smooth phase variations of the modulation profile along a second spatial dimension induce transitions of the edge modes from being localized at one boundary, to a bulk mode, and finally to a localized mode at the opposite boundary. In here, harmonic stiffness modulation profiles are first investigated to illustrate pumping numerically. Square-wave thickness modulations are then employed in the experimental demonstration of the concept. While the majority of studies has so far focused on discrete lattice systems, our results provide a general strategy to achieve topological pumping through continuous property modulations and open new paths towards exploring higher dimensional topological phases exploiting higher dimensions in continuous systems.  

\begin{figure*}[t!]
	\centering
	\begin{minipage}{0.33\textwidth}
		\subfigure[]{
			\includegraphics[width=0.9\textwidth]{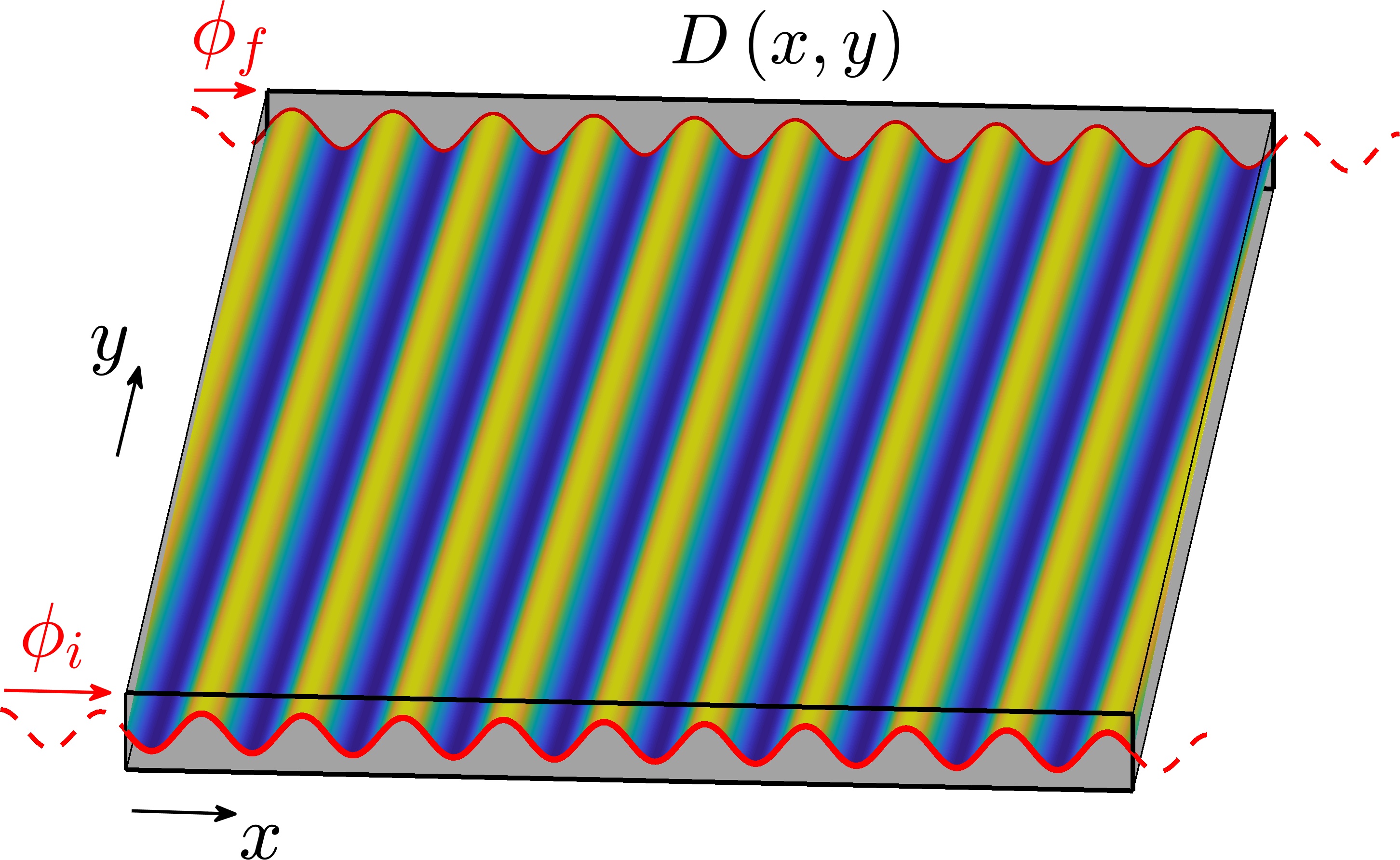}\label{Fig1a}}
		\subfigure[]{
			\includegraphics[width=0.9\textwidth]{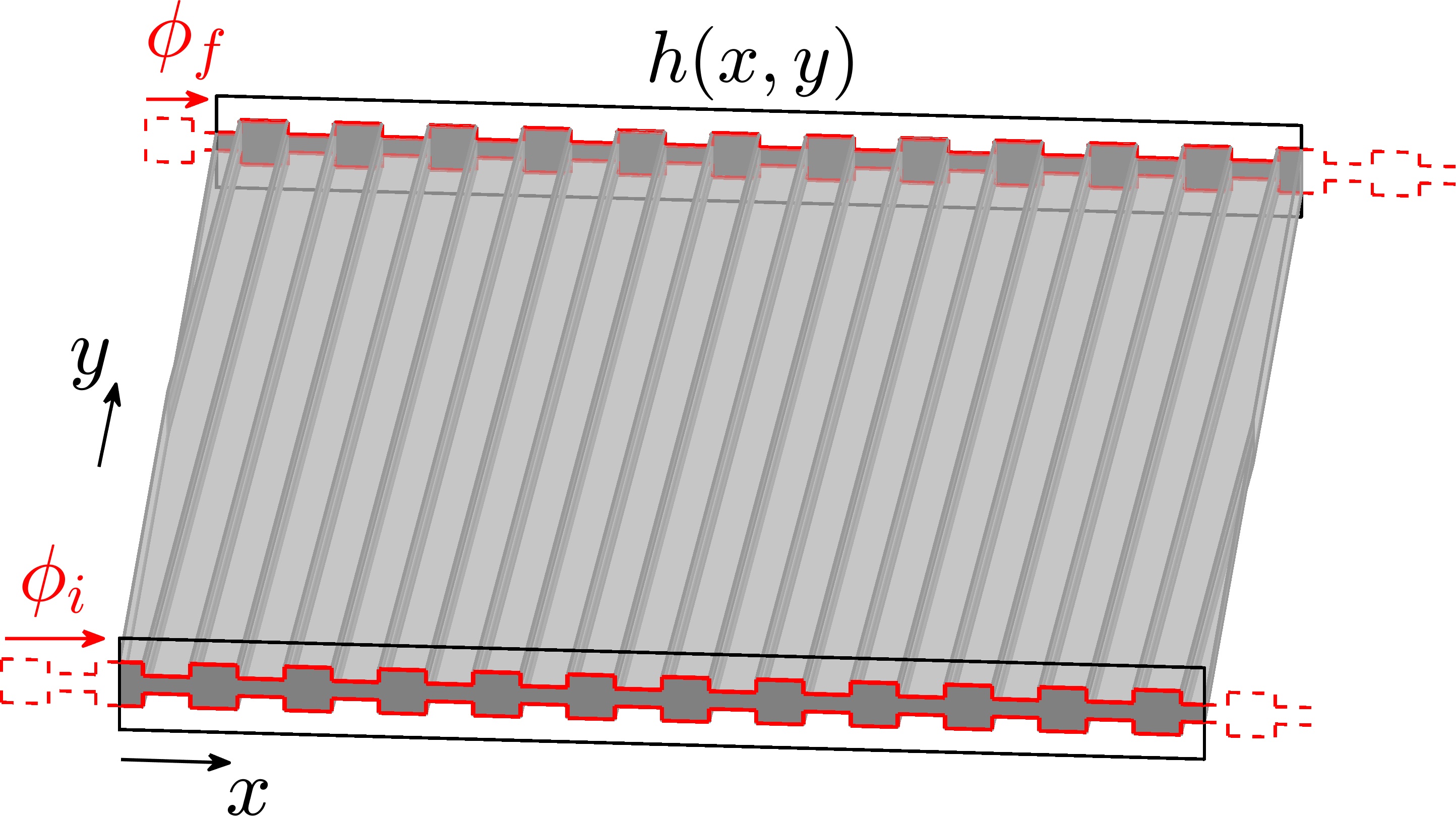}\hspace{-2mm}\label{Fig1b}}
	\end{minipage}\hspace{1mm}
	\begin{minipage}{0.32\textwidth}
		\subfigure[]{
			\includegraphics[width=0.98\textwidth]{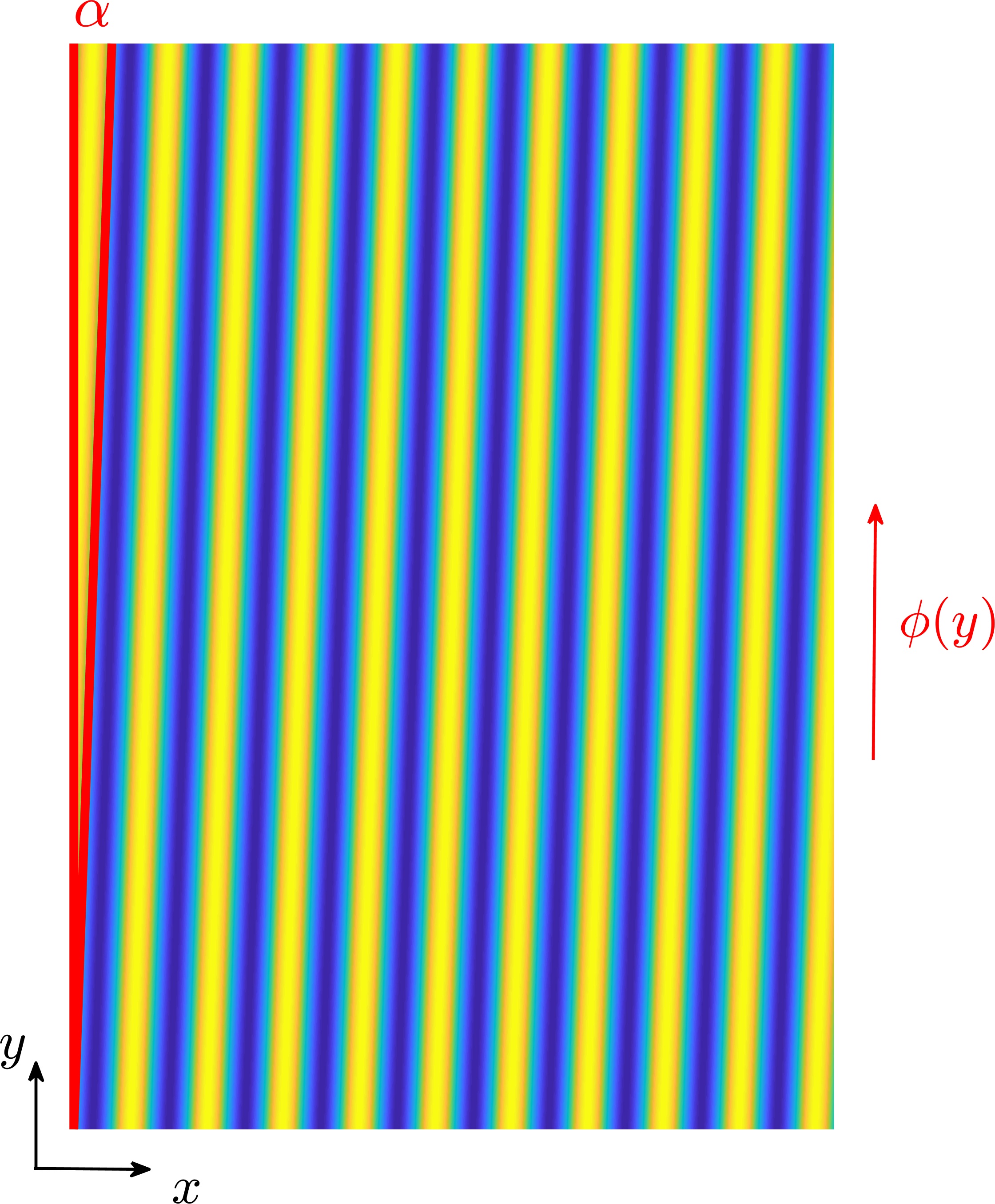}\label{Fig1c}}
	\end{minipage}
	\begin{minipage}{0.33\textwidth}
		\subfigure[]{
			\includegraphics[width=1\textwidth]{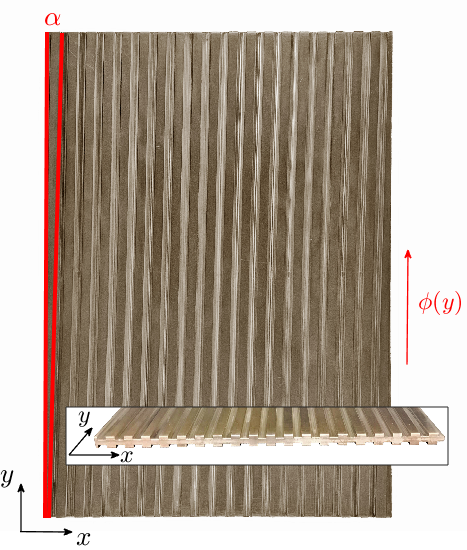}\label{Fig1d}}
	\end{minipage}
	\caption{Stiffness modulations and plate configurations. (a) Plate (shaded gray solid) characterized by a harmonic stiffness modulation $D(x,y)=D_0[1+a_m\cos(\kappa_m x + \phi(y))]$ (colored surface). The schematic illustrates a linear phase change from $\phi_i$ to $\phi_f$. (b) Schematic of plate with square-wave modulation of thickness $h(x,y)=h_0[1+a_m\sign(\cos(\kappa_m x + \phi(y)))]$. The phase also varies linearly from $\phi_i$ to $\phi_f$. (c) Top view of modulation in (a) illustrating the shift of the profile characterized by a tilting angle $\alpha$. (d) Top view and perspective view (inset) of square wave modulated plate employed in experiments. The sample is characterized by parameters $\lambda_m=1.6$ cm, $h_0=4.7$ cm, $a_m=0.38$,  $L_x=31.2$ cm, $L_y=43.7$ cm and phase varying linearly from $\phi_i=0.7\pi$ to $\phi_f=-0.7\pi$.  }
	\label{Fig1}
\end{figure*}

\section{Analaysis of edge states and topological pumping in modulated plates}\label{theorysec}
We consider elastic plates characterized by a bending stiffness which is periodically modulated along the $x$ direction, \textit{i.e.} $D(x,y) = D(x+\lambda_m,y)$, where $\lambda_m$ is the modulation wavelength. Two configurations are investigated in this work (Fig.~\ref{Fig1}). The first one employs a conceptual harmonic stiffness modulation of the form
\begin{equation}\label{eq:mod1}
D(x,y)=D_0[1+a_m\cos(\kappa_m x + \phi(y))]
\end{equation}
where $\kappa_m=2 \pi/\lambda_m$, while $a_m$, and $\phi$ respectively denote amplitude and phase of the modulation. The phase $\phi(y)$ determines the stiffness value $D(0,y)$ at the left boundary of the plate. If smoothly varied along $y$, it produces the tilted modulation profile shown in Figs.~\ref{Fig1}(a,c). This choice follows previous work where sinusoidal modulations define the coupling within continuous waveguides in the context of topological adiabatic pumping~\cite{rosa2019edge}.

The second configuration corresponds to a thickness profile $h(x,y)$ described by a square wave of the form (Fig.~\ref{Fig1}(b)):
\begin{equation}\label{eq:mod2}
h(x,y)=h_0[1+a_m\sign(\cos(\kappa_m x + \phi(y)))]
\end{equation}
which produces a periodic modulation of the plate bending stiffness according to the expression $D(x,y)=Eh(x,y)^3/(12(1-\nu^2))$, where $E,\nu$ are respectively the Young's modulus, and the Poisson's ratio of the plate material. This choice is driven by fabrication considerations in the experimental activities of this work. 

The effects of the harmonic stiffness modulation (Eqn.~\eqref{eq:mod1}) are investigated analytically by considering Kirchhoff-Love's plate theory~\cite{graff2012wave}. According to the theory, the harmonic motion at frequency $\omega$, $w(x,y,\omega)$, in the direction perpendicular to the plate plane $x,y$ is governed by the following equation of motion:
\begin{equation}\label{Eq.plate}
\begin{split}
\left[D \left( w_{,xx}+\nu w_{,yy}\right) \right]_{,xx}+2\left[ (1-\nu)Dw_{,xy} \right]_{,xy} \\
+\left[D \left( w_{,yy}+\nu w_{,xx} \right) \right]_{,yy} = \omega^2 m w,
\end{split}
\end{equation}
where $()_{,q}$ denotes a partial derivative with respect to $q$, and $m=\rho h$ is the mass density. We investigate the dispersion properties of the plate $\omega=\omega(\kappa_x, \kappa_y,\phi)$, where the phase modulation $\phi$ is explicitly denoted as a free parameter. To this end, we impose plane wave solutions $w(x,y)=\textup{w}(x) e^{j \kappa_y y}$, where $\textup{w}(x)=\sum_{n} \hat{w}_n e^{j (\kappa_x + n \kappa_m) x}$, $n=-N,..,+N$ reflects the $x$-wise periodicity of the plate. Application of the Plane Wave Expansion Method (PWEM) (see Supplemental Material (SM)~\cite{SM}), leads to an eigenvalue problem in the form:
\begin{equation}\label{eigprob}
\bm{K}(\kappa_x,\kappa_y,\phi)\hat{\bm{w}}=m\omega^2 \hat{\bm{w}},
\end{equation}
where $\bm{K}$ is the $N \times N$ stiffness matrix and $\hat{\bm{w}}=\{ \hat{w}_{-N}, ..., \hat{w}_{N} \}^T$. Solution of the eigenvalue problem in Eqn.\eqref{eigprob} yields the dispersion properties, described in terms of eigenvalues $\omega_i$ and associate wave modes $\textup{w}_i$ defined by the components of the eigenvector $\hat{\bm{w}}_i$.

\begin{figure*}[!ht]
	\centering
	\subfigure[]{\includegraphics[height=0.265\textwidth]{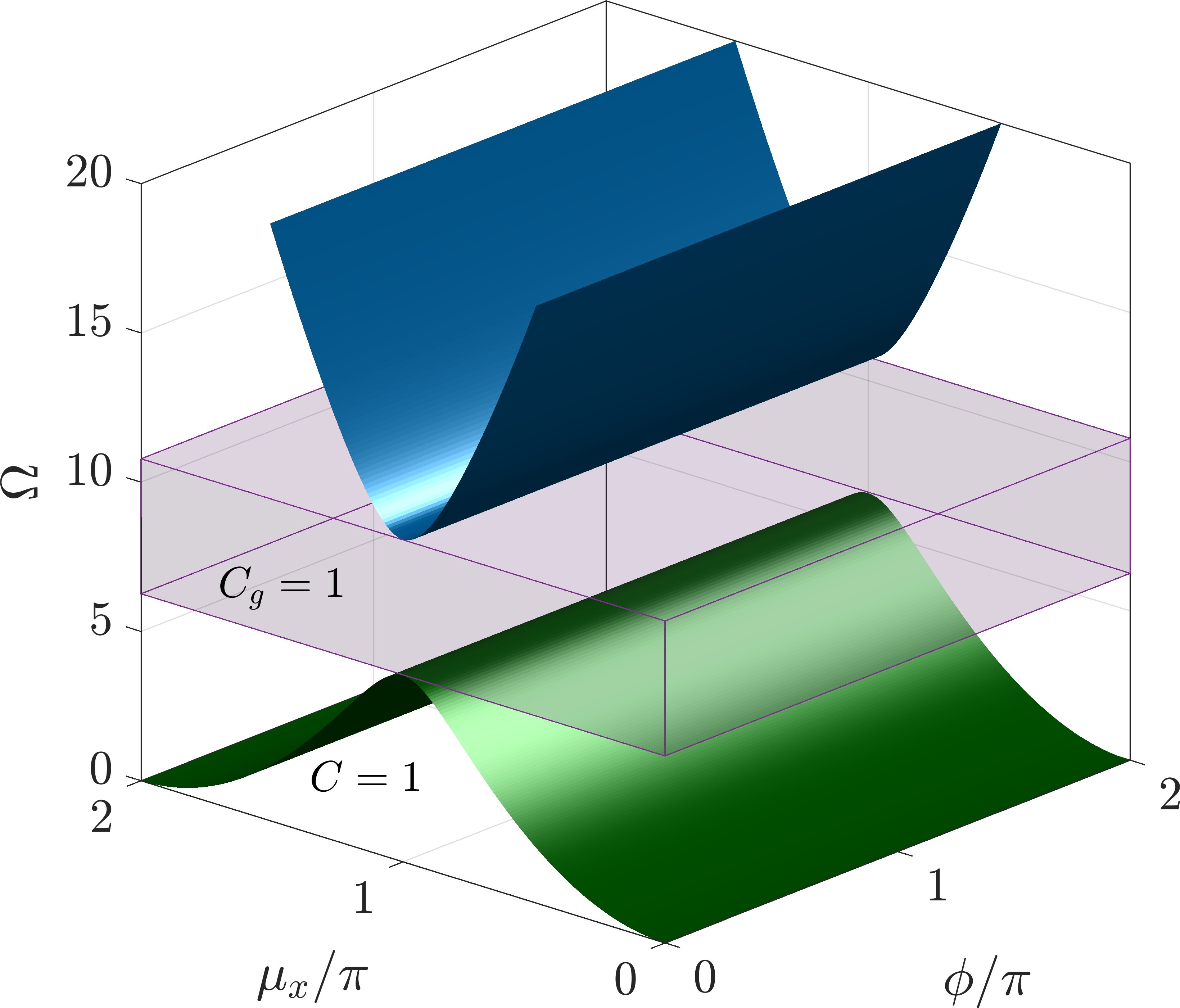}\label{Fig2a}}
	\subfigure[]{\includegraphics[height=0.265\textwidth]{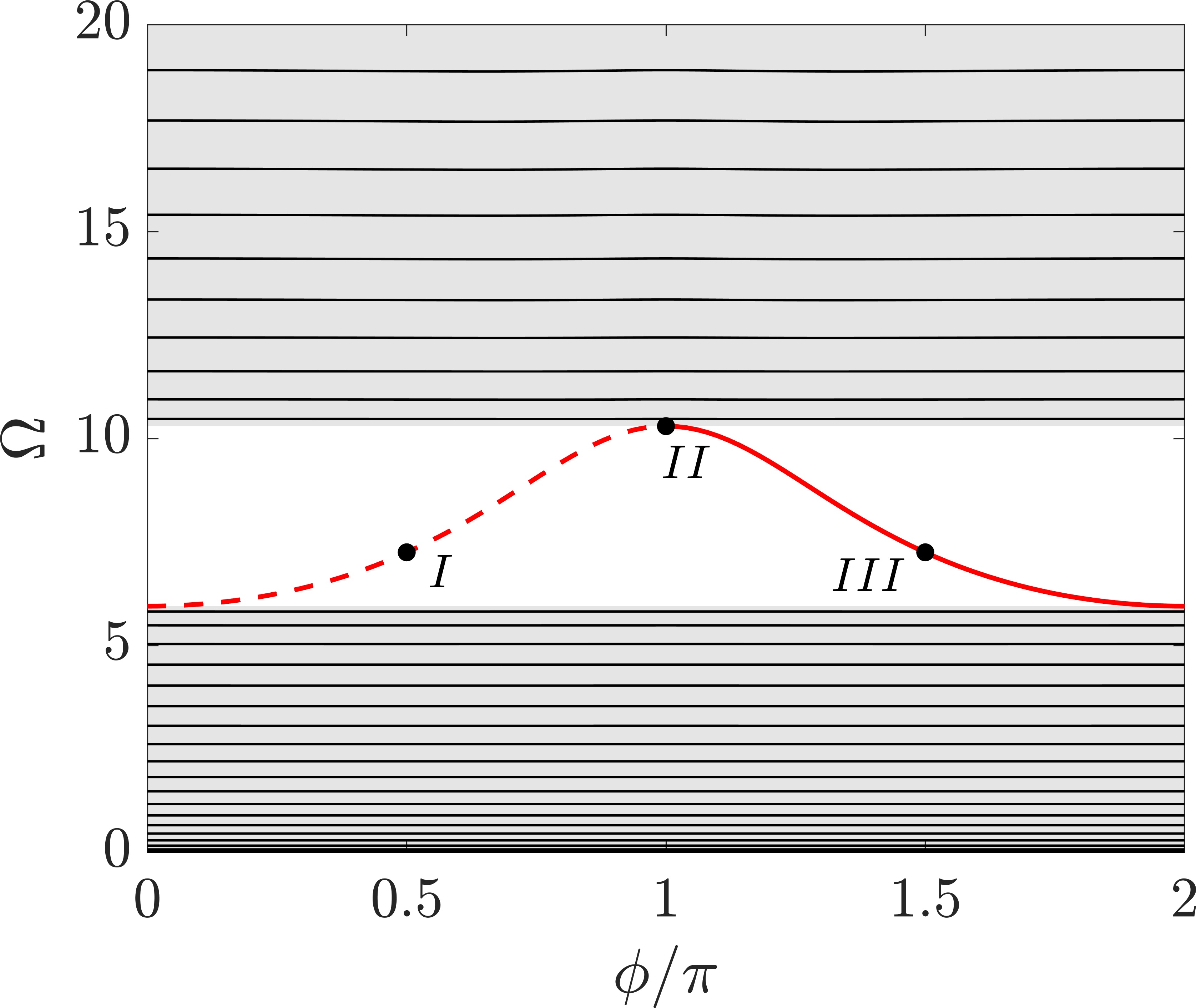}\label{Fig2b}}\hspace{3mm}
	\subfigure[]{\includegraphics[height=0.265\textwidth]{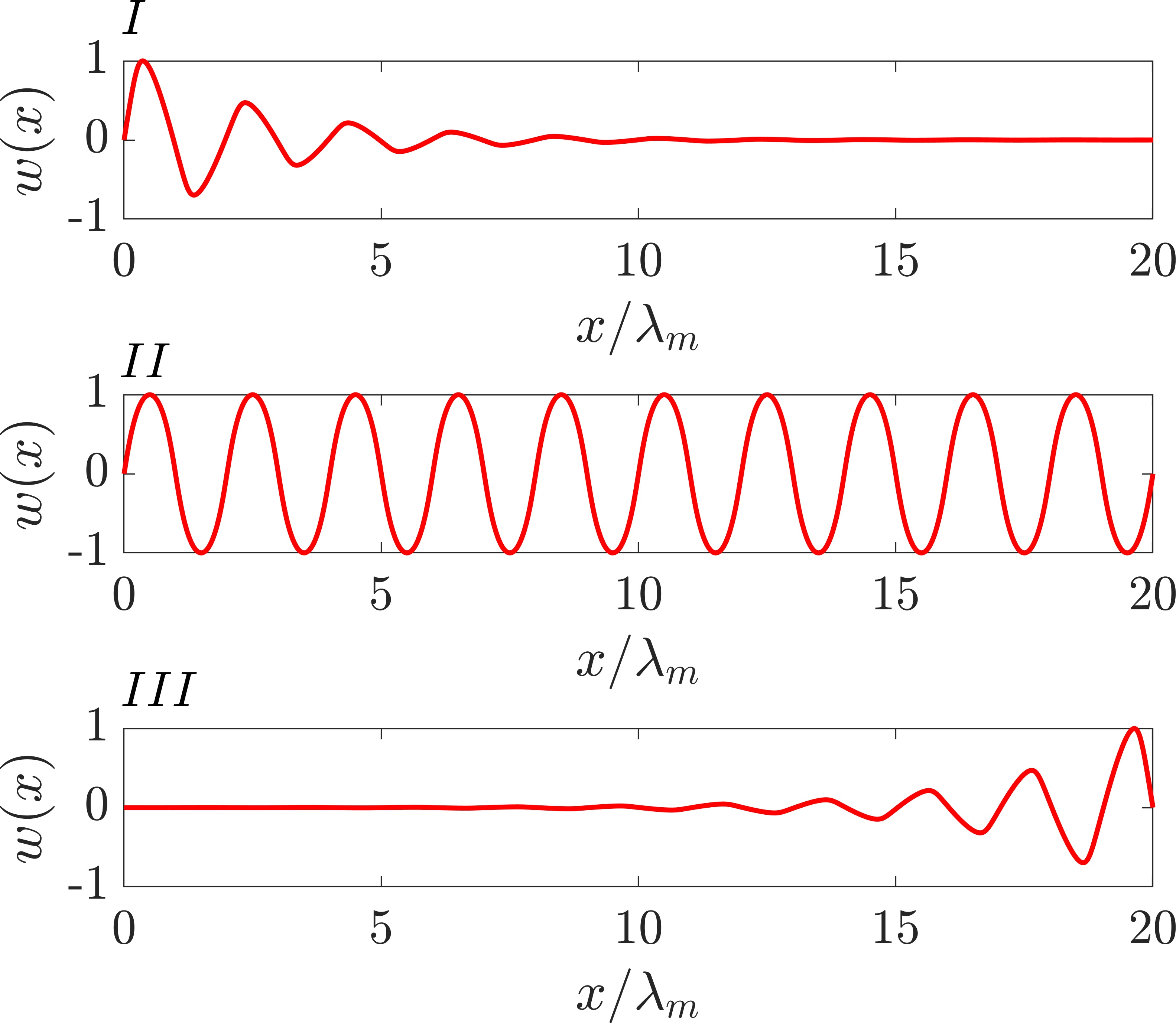}\label{Fig2c}}
	\subfigure[]{\includegraphics[height=0.265\textwidth]{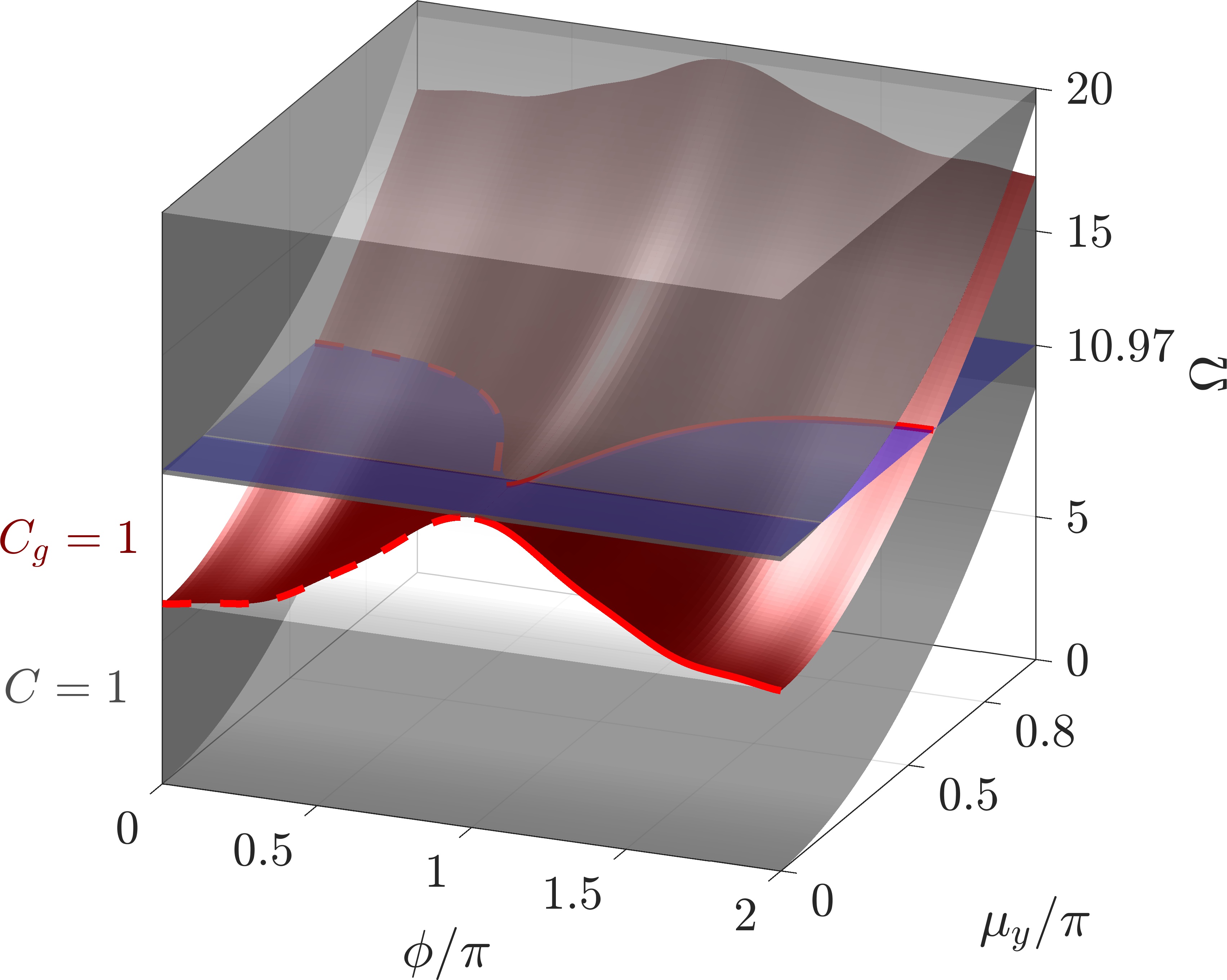}\label{Fig2d}}
	\subfigure[]{\includegraphics[height=0.265\textwidth]{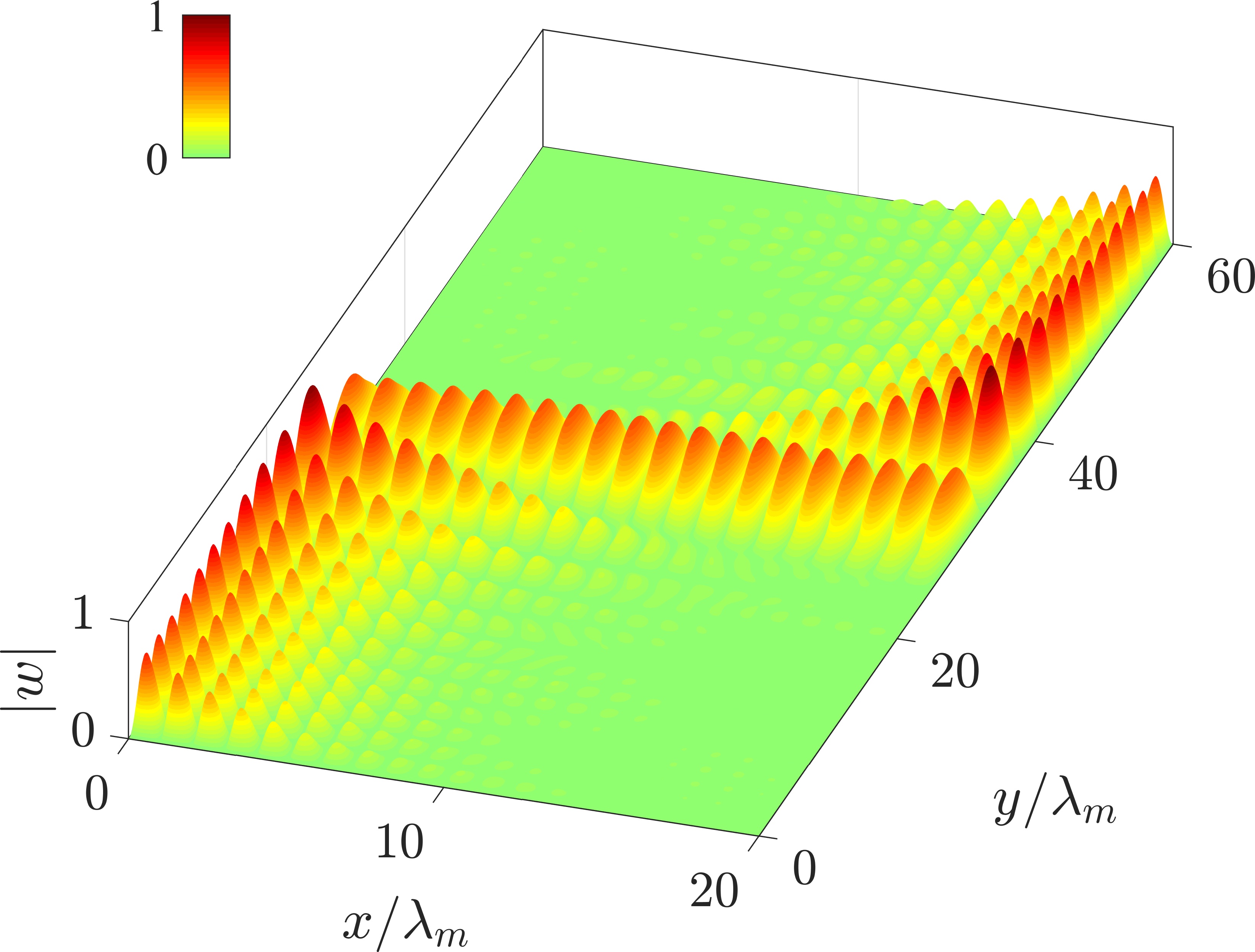}\label{Fig2e}}
	\subfigure[]{\includegraphics[height=0.265\textwidth]{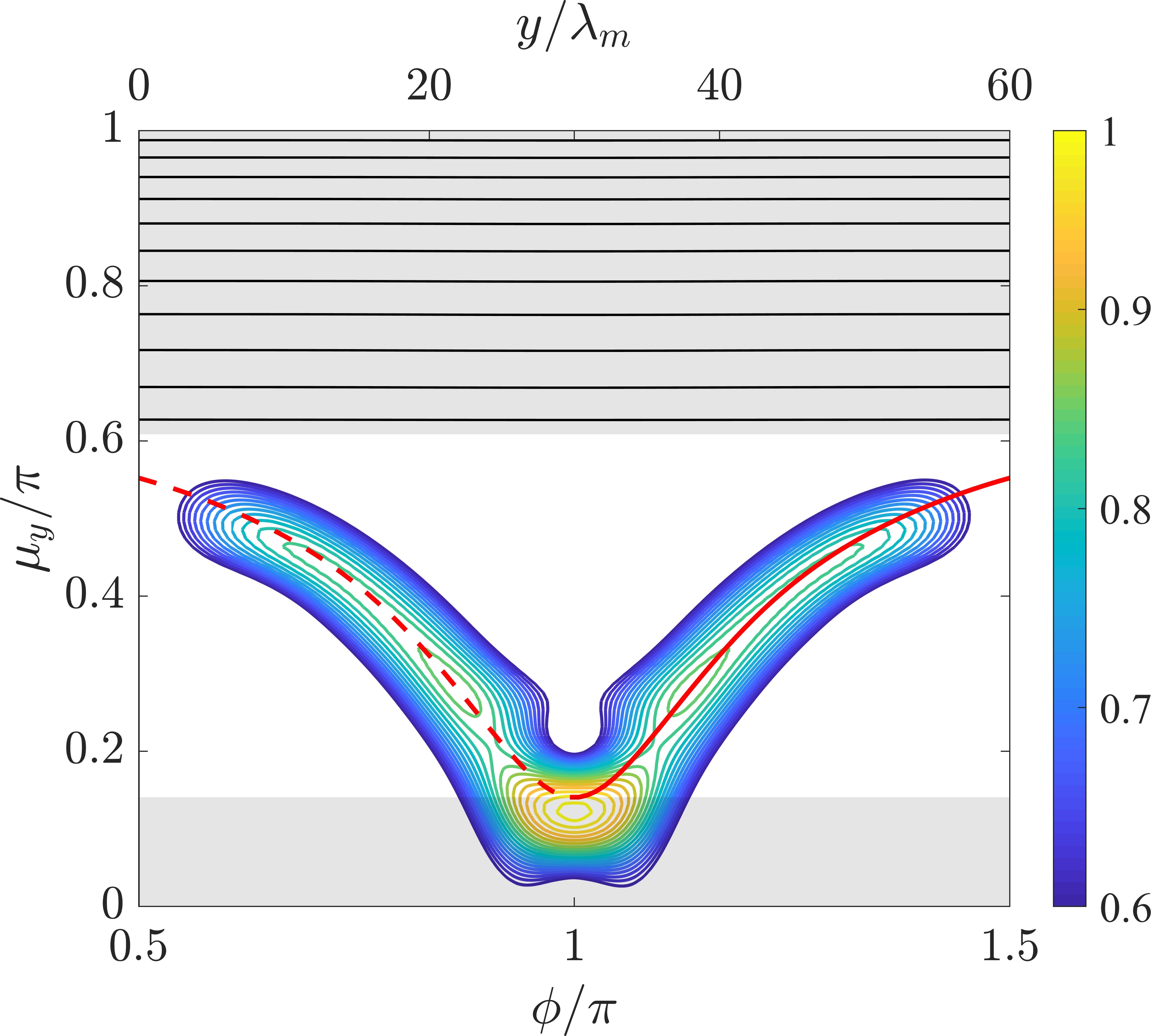}\label{Fig2f}}
	\caption{Dispersion properties and topological pumping for elastic plate with harmonic stiffness modulation $D(x,y)=D_0[1+a_m\cos(\kappa_m x + \phi(y))]$. (a) Dispersion surfaces $\Omega(\mu_x,\phi)$ for $\mu_y=0$ showing two bands separated by a gap, with information on Chern numbers and gap label. (b) Frequency spectrum of finite plate with $L_x=20\lambda_m$ and $\mu_y=0$ as a function of $\phi$: black lines corresponding to finite structure modes are superimposed to the bulk bands (shaded gray regions), while an edge mode (red line) spans the gap. (c) Representative modes for left-localized, bulk and right-localized modes, corresponding to the points marked in (b). (d) Variation of the finite plate spectrum in (b) as a function of $\mu_y$. The red surface represents the dispersion of the edge state, while shaded gray volumes are the bulk bands. The red line at $\mu_y=0$ highlights the transition of the edge state from left-localized (dashed) to right localized (solid) that occurs at $\phi=\pi$. (e) Steady-state response $|w(x,y)|$ of the modulated plate at frequency $\Omega=10.97$, where the associated colormap also represents normalized displacement. Topological pumping occurs through a transition of the edge state from left-localized ($\phi_i=0.5\pi$) to right localized ($\phi_f=1.5\pi$) due to the phase modulation $\phi(y)=\phi_i \to \phi_f$. (c) Cross section of dispersion diagram in (d) at frequency $\Omega=10.97$ as a function of $\phi \in [0.5\pi, 1.5\pi]$. Black and red lines respectively denote bulk and edge modes of the finite plate. The contours represent the spectrogram of the displacement field $|\hat{w}(y,\mu_y)|$, revealing that pumping occurs through a transition along the wavenumber branch of the edge state.}
	\label{Fig2}
\end{figure*}

We begin our study by evaluating dispersion along $x$, for assigned values of $\kappa_y$. Results for $\kappa_y=0$, shown in Fig.~\ref{Fig2a}, correspond to letting $()_{,y}=0$ in Eqn.~\eqref{Eq.plate}, which yields an expression akin to the equation governing the transverse motion of 1D elastic Euler Bernoulli beams~\cite{meirovitch1975elements}. In Fig.~\ref{Fig2a}, $\mu_x=\kappa_x \lambda_m$, while $\Omega=\omega/\omega_0$ is a non-dimensional frequency, with $\omega_0=\kappa_m^2 \sqrt{D_0/m}$. The results, obtained for $a_m=0.8$, effectively correspond to the dispersion characteristics of a family of 1D, decoupled elastic beams characterized by stiffness modulations that differ by the phase parameter $\phi$. The dispersion eigenvalues feature two bands separated by a gap that remains constant with $\phi$. Indeed, continuous shifts of the stiffness along $x$ can be interpreted as a translation of the $\lambda_m$-periodic unit cell along $x$, which does not affect the eigen-frequencies. However, these shifts do affect the eigenvectors, as it is revealed by the analysis of the topology of the bands. Such analysis relies on the evaluation of band's Chern number in the $(\mu_x,\phi) \in \mathbb{T}^2 = [0,2\pi]\times [0,2\pi]$ space~\cite{hatsugai1993chern,rosa2019edge}, which is given by
\begin{equation}\label{Cherneq}
C = \dfrac{1}{2 \pi i} \int_{\mathcal{D}} \nabla \times(\textup{w}_i^*\cdot \nabla \textup{w}_i) \; d\mathcal{D},
\end{equation}
where $\mathcal{D}=\mathbb{T}^2$, $\nabla = (\partial/\partial \mu_x)\bm{e}_{\mu_x} +  (\partial/\partial \phi)\bm{e}_{\phi} $ and $()^*$ denotes a complex conjugate. The Chern number is evaluated numerically over a discretized $(\mu_x,\phi)$ space according to the procedure described in~\cite{fukui2005chern}, which gives the label assigned to the first band in Fig.~\ref{Fig2a}. A label for a gap $r$ is then assigned by computing the algebraic sum of the Chern numbers of the bands below it~\cite{hatsugai1993chern,rosa2019edge}, \textit{i.e.} $C_g^{(r)}=\sum_{n=1}^{r}C_n$, which yields $C_g=1$ for the gap considered in Fig.~\ref{Fig2a}. In finite structures, a non-zero gap label signals the presence of topological edge states spanning the associated gap as a result of a parameter sweep. The existence of an edge state as $\phi$ varies in the $[0,\,\, 2\pi]$ range is verified by computing the spectral properties of a plate bounded along the $x$ direction, which are evaluated by constructing an eigenvalue problem similar to that of Eqn.~\eqref{eigprob}, where a solution of the kind 
$w(x,y)=e^{j \kappa_y y} \sum_{n} \hat{w}_n \sin(\frac{n \pi x}{L_x}), \,\, n=1,..,N$
is imposed to satisfy the conditions at the plate $x$-boundaries, \emph{i.e.} $w(x=0,L_x;y)=w_{,xx}(x=0,L_x;y)=0$~\cite{SM}. Figure~\ref{Fig2b} shows the modes of a finite plate of length $L_x=20\lambda_m$ as a function of $\phi$, for $\kappa_y=0$. The modes (black solid lines) belonging to the bulk bands, which are shown for reference as the shaded gray regions, do not vary as a function of $\phi$. An additional mode (red line) traversing the gap and varying with $\phi$ corresponds to a topological edge state localized at either the left or right boundary depending on the value of $\phi$, with the right (left) localization of the mode being denoted by the solid (dashed) red line. The transition of the edge state with variations of $\phi$ is related to the gap label $C_g=1$. In particular, its absolute value $|C_g|=1$ indicates that the edge state traverses the gap once for $\phi \in [0,2\pi]$, while its positive sign relates to a left-to-right transition that occurs when the branch of the edge state touches the upper boundary of the gap at $\phi=\pi$. Representative left-localized (point $I$) and right-localized modes (point $III$) are displayed in Fig.~\ref{Fig2c}, along with the mode extending to the bulk in correspondence to the branch touching the bulk band (point $II$). These observations are in agreement with the behavior of edge states and their correspondence to the gap labels in discrete lattices~\cite{hatsugai1993chern,rosa2019edge}. 

Next, we discuss the dispersion properties for values $\kappa_y\neq0$. These values do not affect the structure of the eigenvalue problem in Eqn.~\eqref{eigprob}, and only introduce a frequency shift. This is illustrated in Fig.~\ref{Fig2d} which displays the dispersion of a modulated plate with $a_m=0.8$. The two bands are denoted by the shaded gray volumes, which are spanned for $\mu_x\in[0,\,\, \pi]$. Their variation in terms of $\mu_y=\kappa_y\lambda_m$ and $\phi$ illustrates the presence of a separating gap at frequencies that increase monotonically with $\mu_y$. The modes of a finite plate with $L_x=20\lambda_m$ populate these bulk bands, here omitted for simplicity, and also include a mode spanning the gap, which is represented by the red surface in the figure. The red lines superimposed at $\mu_y=0$ illustrate the transition experienced by the edge state as in Fig.~\ref{Fig2b}, which now occurs as a function of $\mu_y$ along the entire surface of the edge state. To further confirm the topological properties, the Chern number and gap label as previously defined are evaluated as a function of $\mu_y$ in discretized $(\mu_x,\phi) \in \mathbb{T}^2 = [0,2\pi]\times [0,2\pi]$ spaces, which yields the labels in Fig.~\ref{Fig2d}. This result is expected from values $\mu_y\neq0$ not affecting the dispersion topology. 

The transitions of the edge states can be exploited to implement a topological pump that employs an adiabatic (slow) variation of $\phi$ along a second dimension~\cite{kraus2012topological,rosa2019edge,nassar2018quantization,grinberg2019robust}. For a finite plate of length $L_y$ , we consider a smooth, linear phase modulation of the kind $\phi(y)=\phi_i\left(1-\frac{y}{L_y}\right) + \phi_f\frac{y}{L_y}$ (Fig.~\ref{Fig1a}). A top view of a representative harmonic stiffness modulation is displayed in Fig.~\ref{Fig1c}, where a positive tilting angle $\alpha=\tan^{-1}\left(-\left(\phi_f-\phi_i\right)/\left(\kappa_m L_y\right)\right)$ resulting from a choice with $\phi_i>\phi_f$ is illustrated. We first demonstrate topological pumping numerically by considering a plate with $L_y=3L_x$ and phase variation with $\phi_i=0.5\pi$ and $\phi_f=1.5\pi$. These values cause the edge states to transition from the left boundary to the right boundary. To verify this, we compute the forced response of the plate when harmonically excited by a distributed force per unit area $q(x,y,t)=f(x)\delta(y-y_{e})e^{i\omega t}$. The force is applied near the bottom boundary ($y_{e}=\lambda_m/2$), and has a spatial distribution $f(x)$ that corresponds to the left-localized edge state obtained for $\phi_i=0.5\pi$ (see Mode $I$ in Fig.~\ref{Fig1c}). This favors the excitation of the desired topological mode, while minimizing the contribution from bulk modes co-existing at the same frequency. The response of the plate is evaluated through a Galerkin~\cite{meirovitch1975elements} approximation of the displacement field $w(x,y)$, similar to that employed to obtain the modes of the finite plate (see details in SM~\cite{SM}). The response of the plate for an excitation frequency of $\Omega=10.97$ (Fig.~\ref{Fig2e}) consists of a topological pump whereby energy is transferred from the bottom left boundary to the upper right boundary of the plate via an edge state transition. Additional examples are reported in the Supplementary Material (SM)~\cite{SM}. The topological pumping results from an adiabatic evolution along the wavenumber branch of the edge state at a given frequency~\cite{rosa2019edge}, which is illustrated for the pump of Fig.~\ref{Fig2e} by considering a cross-section of the dispersion diagram at frequency $\Omega=10.97$ (blue plane in Fig.~\ref{Fig2d}). The results in Fig.~\ref{Fig2f} are displayed for $\phi \in [0.5\pi, 1.5\pi]$, which is the interval considered for the phase modulation $\phi(y)$. Shaded gray areas correspond to the intersection between the blue plane and the bulk bands (shaded gray volumes) of Fig.~\ref{Fig2d}, and represent the dispersion bands $\mu_y(\Omega=10.97,\phi)$ occupied for $\mu_x \in \{0, \pi \}$. These bands are populated by modes of the finite plate with $L_x=20\lambda_m$ (solid black lines), while the intersection between the blue plane and the red surface in Fig.~\ref{Fig2d} defines the edge state (red line) that spans the gap as a function of $\phi$. The previously described forcing profile selectively excites the left-localized edge state (for $\phi_i=0.5\pi$) at the bottom boundary of the plate, while a smooth phase modulation $\phi(y)=\phi_i \to \phi_f$ drives the left-to-right transition of the edge state along $y$, which occurs along the branch defined by the red lines in Fig.~\ref{Fig2f}. We verify such transition by computing 2D Fourier Transforms (FT) while performing an appropriate windowing of the displacement field to capture wavenumber changes along $y$. The procedure consists on pre-multiplying the displacement field $\textup{w}(x,y)$ by a Gaussian window centered at $y=y_0$, \textit{i.e.} $G(x,y)=e^{-(y-y_0)^2/2c^2}$, where $c$ is a parameter controlling the Gaussian's width. A FT operation then quantifies the displacement field in reciprocal space $\hat{w}(y_0,\mu_x,\mu_y)$ around the location $y=y_0$. The dependence of $\mu_x$ is then eliminated by taking the $L^1$ norm along $\mu_x$, which produces $\hat{w}(y,\mu_y)$. The corresponding spectrogram, obtained for $c=0.07$, is displayed in the form of contour plots in Fig.~\ref{Fig2f}, where the colors represent the normalized magnitude of the displacement field $|\hat{w}(y,\mu_y)|$. The procedure confirms that energy remains concentrated on the wavenumber branch of the edge state according to the modulation $\phi(y)$, which characterizes the topological pump displayed in Fig.~\ref{Fig2e}.

\begin{figure*}[t!]
	\centering
	\subfigure[]{\includegraphics[height=0.35\textwidth]{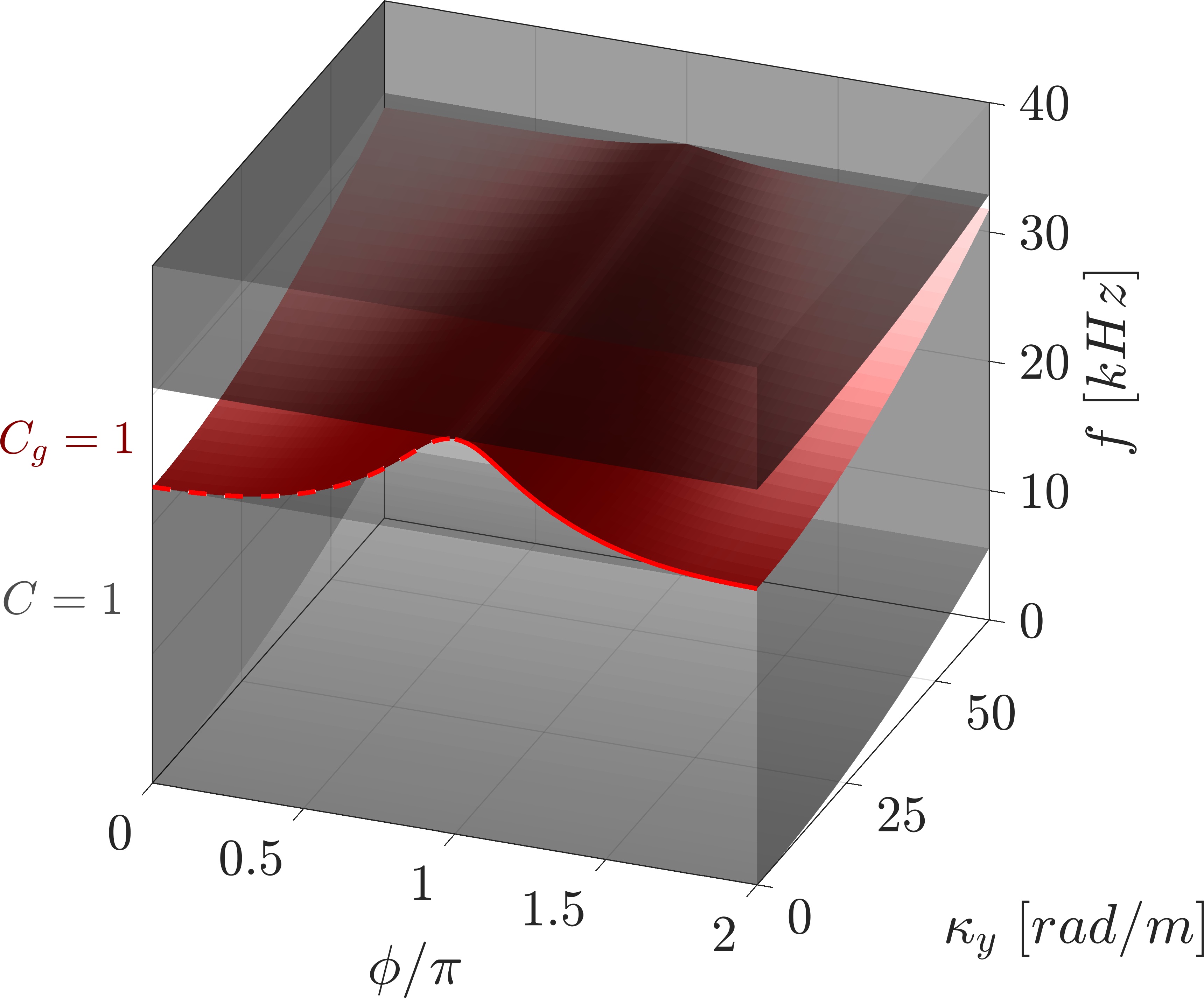}\label{Fig3a}}\hspace{2mm}
	\subfigure[]{\includegraphics[height=0.35\textwidth]{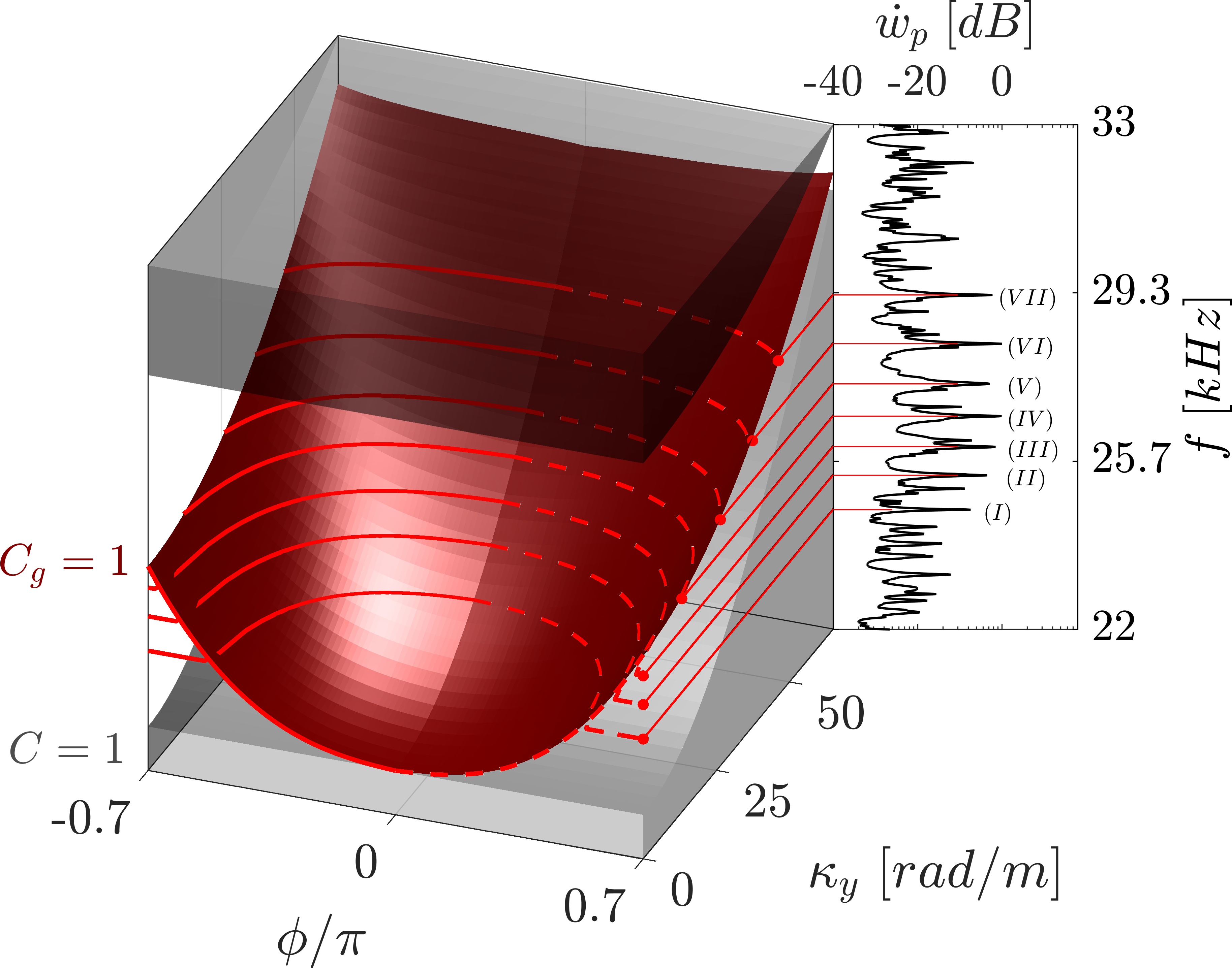}\label{Fig3b}}
	\subfigure[]{\includegraphics[width=0.31\textwidth]{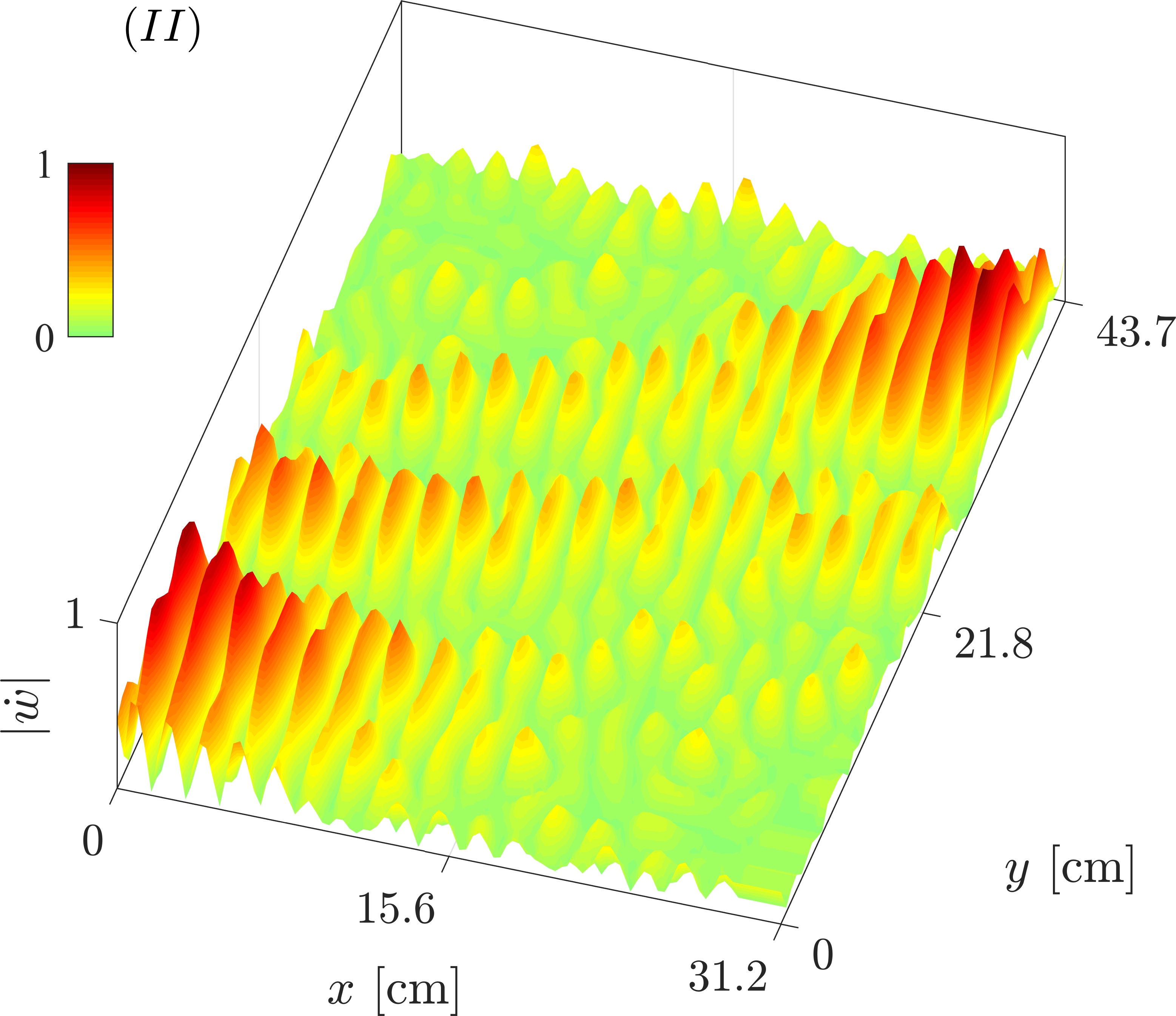}}
	\subfigure[]{\includegraphics[width=0.31\textwidth]{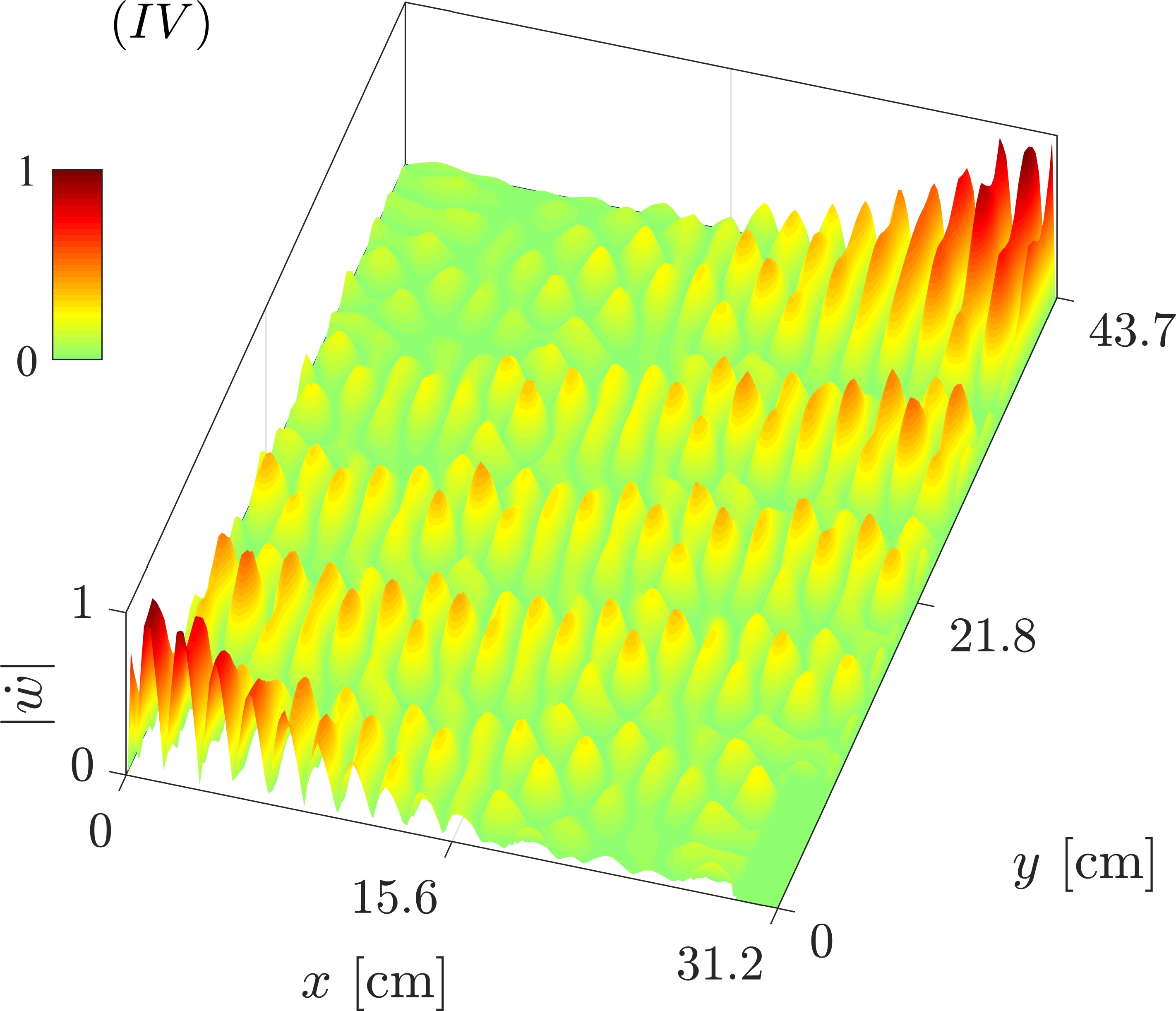}}
	\subfigure[]{\includegraphics[width=0.31\textwidth]{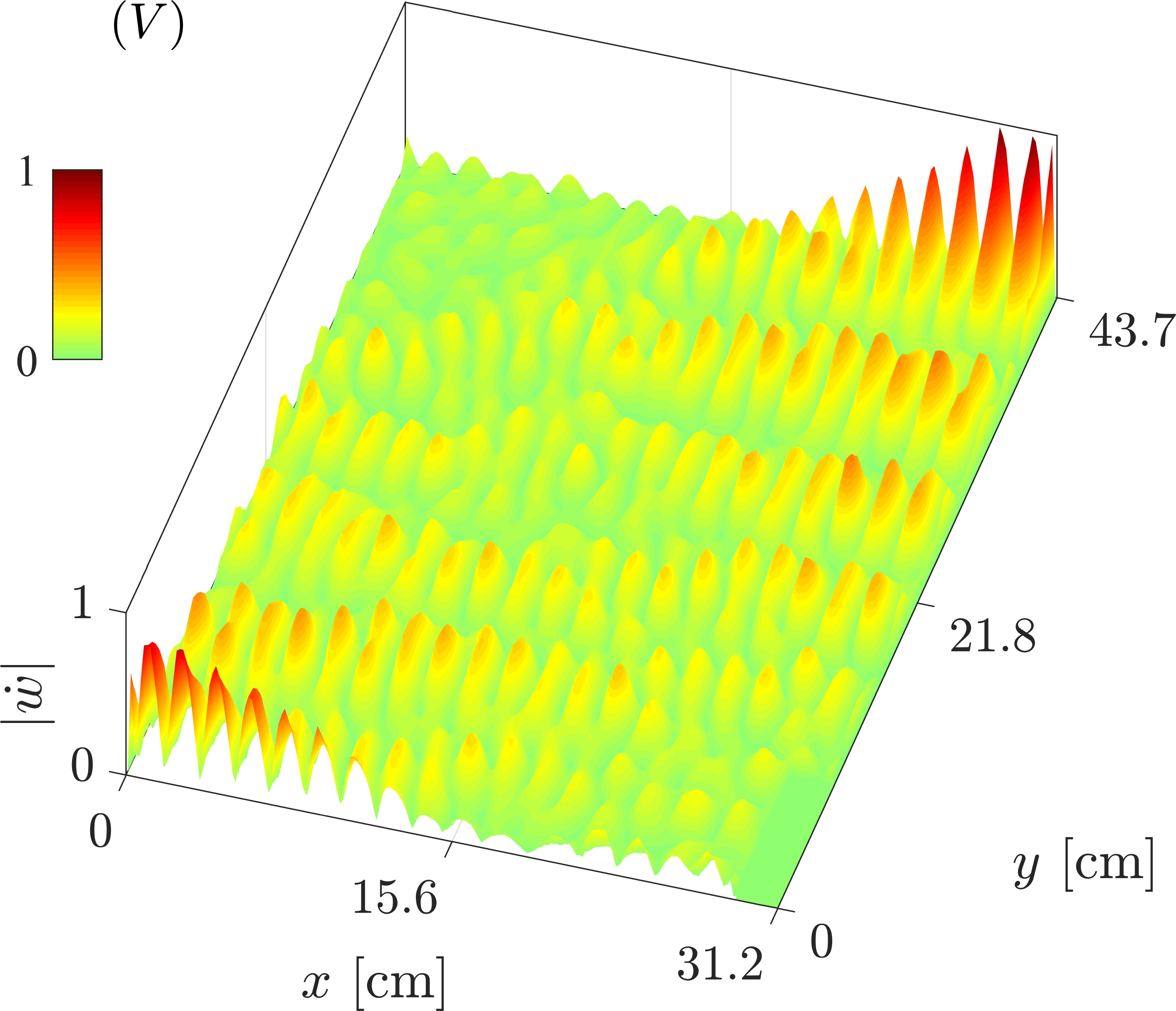}}
	\subfigure[]{\includegraphics[width=0.31\textwidth]{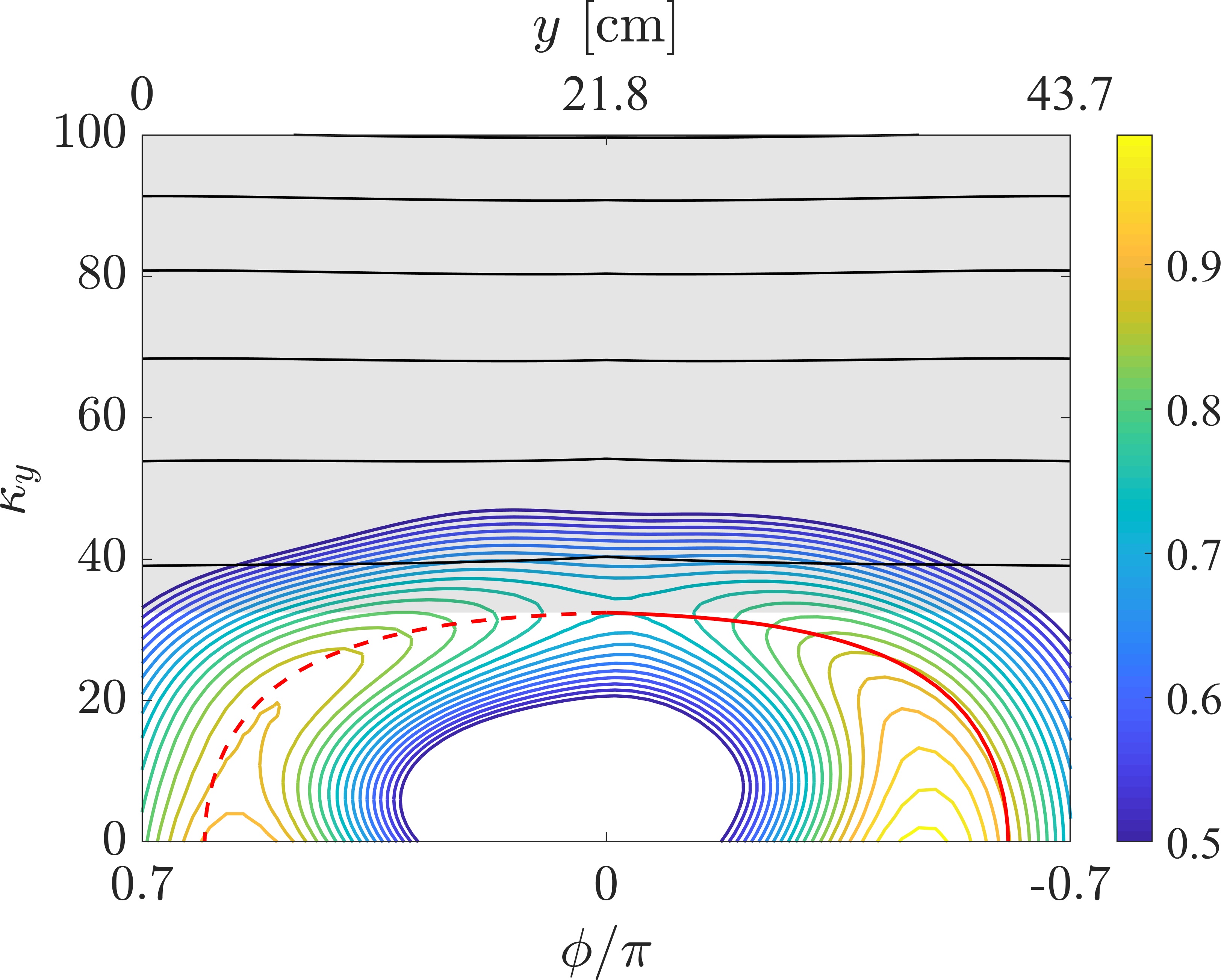}}
	\subfigure[]{\includegraphics[width=0.31\textwidth]{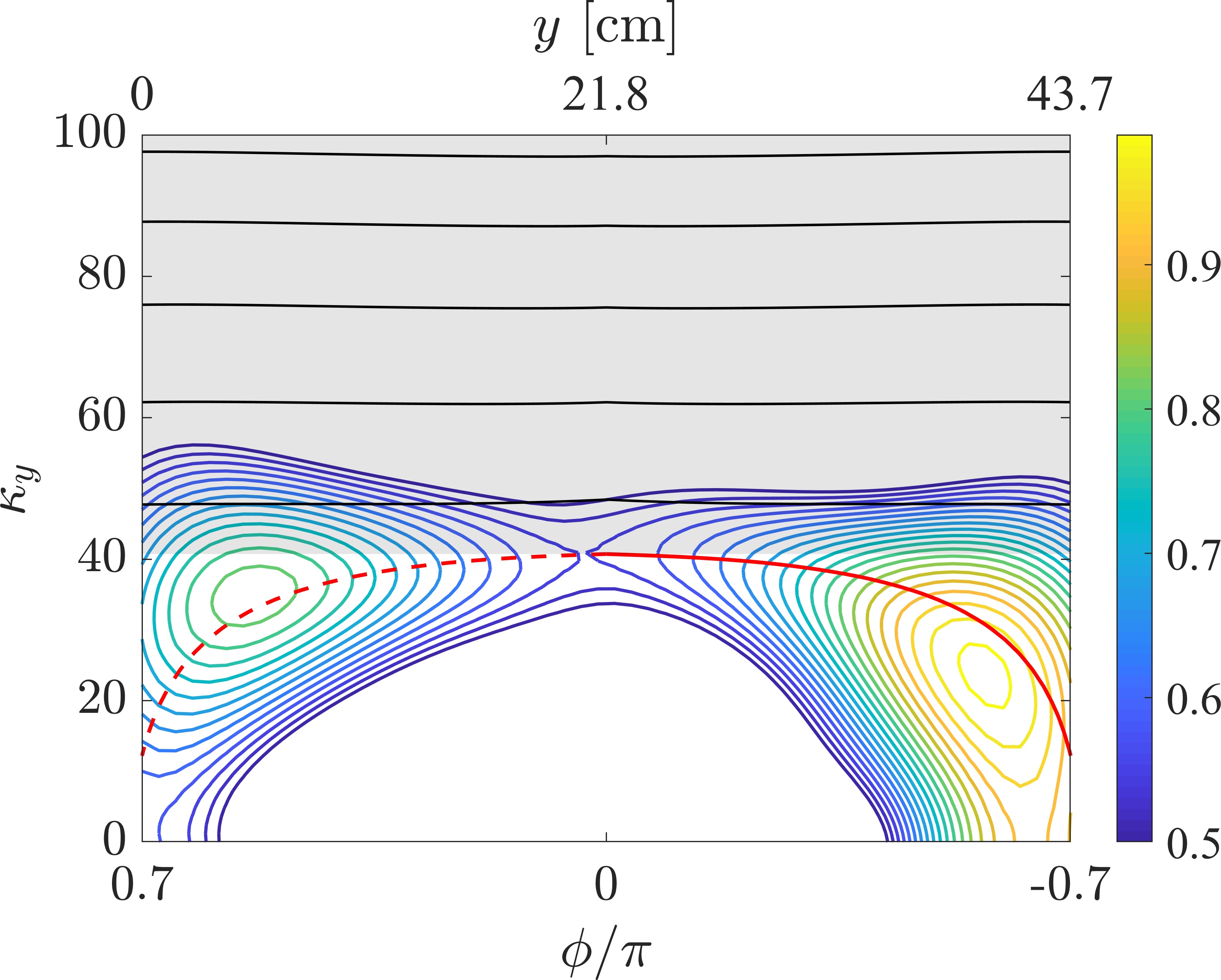}}
	\subfigure[]{\includegraphics[width=0.31\textwidth]{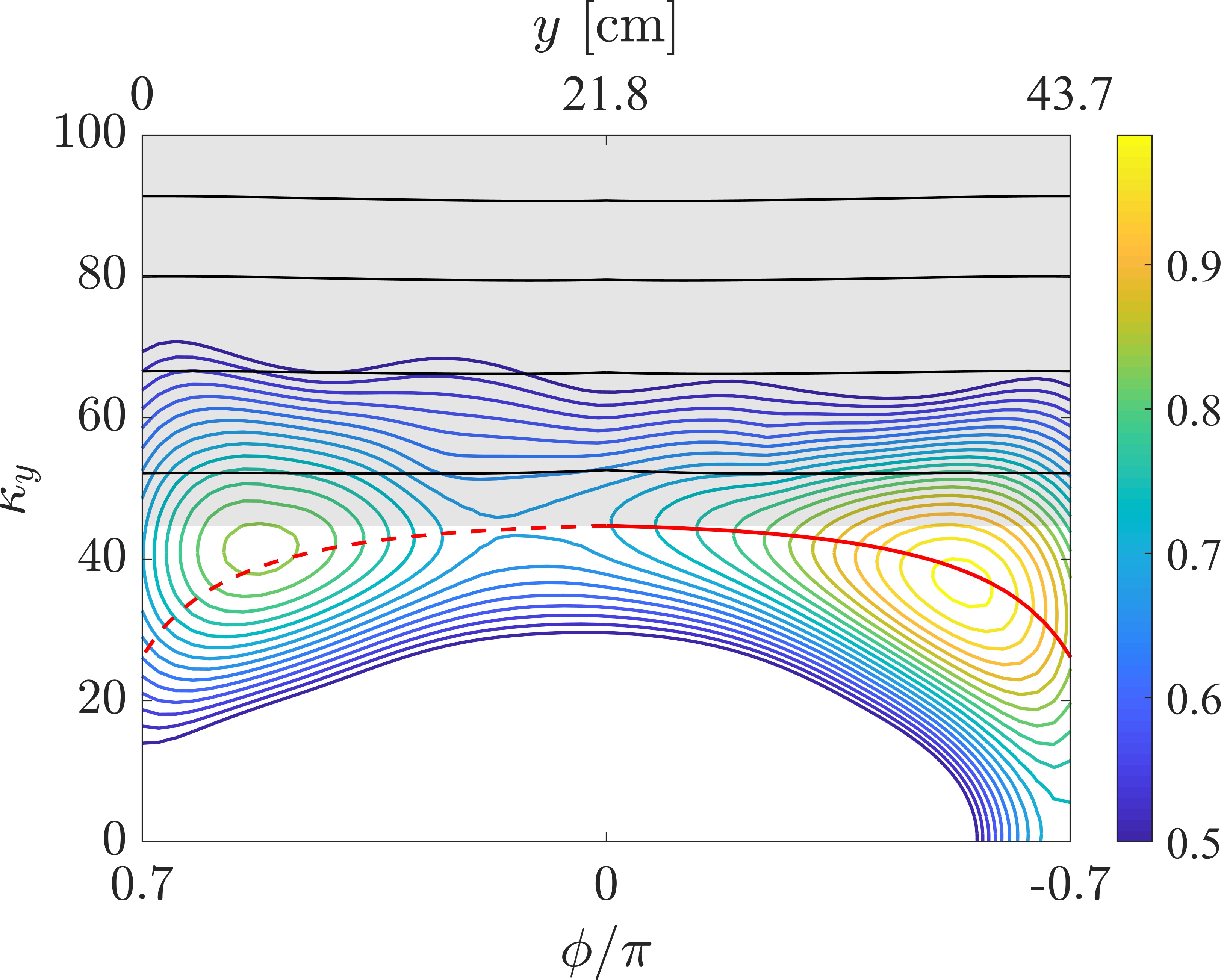}}
	\caption{Dispersion properties and experimental observation of topological pumping in plate with square-wave modulation of the thickness $h(x,y)=h_0[1+a_m\sign(\cos(\kappa_m x + \phi(y)))]$. (a) Dispersion properties and edge state (red surface) for a finite plate with $L_x=31.2$cm and free-free boundary conditions along $x$. Bulk bands are represented by shaded gray volumes, along with Chern number and gap labels information. (b) Detail of dispersion for $\phi \in [-0.7\pi, 0.7\pi]$, corresponding to the interval considered for the phase modulation of thickness. The experimentally measured frequency response spatially averaged over the plate surface is displayed alongside the dispersion. Each resonant peak defines a wavenumber branch highlighted by dashed and solid red lines in the dispersion surface, along which topological pumping occurs at the corresponding frequency. (c-e) Experimentally measured velocity field $|\dot{w}(x,y)|$ for selected resonant peaks $II$, $IV$ and $V$. The transitions from localization at the left-boundary to localization at the right-boundary that characterize topological pumping are quantified by the spectrograms displayed in (f-h), which confirm that energy is concentrated around the wavenumber branches of the edge states. }
	\label{Fig3}
\end{figure*}

\section{Experimental observation of topological pumping in square-modulated plate}\label{experimentsec}

Topological pumping is experimentally demonstrated in a plate with square-wave thickness modulation $h(x,y)=h_0[1+a_m\sign(\cos(\kappa_m x + \phi(y)))]$ with modulation parameters $\lambda_m=1.6$ cm, $h_0=4.7$ cm, $a_m=0.38$, $\phi_i=0.7\pi$ and $\phi_f=-0.7\pi$ (Fig.~\ref{Fig1b}). The plate is rectangular of dimensions $31.2$ cm $\times$ $43.7$ cm, and it is made of aluminum (Fig.~\ref{Fig1d}). A linear phase modulation produces the tilted thickness profile in Fig.~\ref{Fig1d}.

The plate dispersion properties are computed using a Finite Element (FE) model implemented in the COMSOL Multiphysics environment~\cite{SM}. The computations provide information on all wave modes, whose polarizations are tracked by computing a polarization factor~\cite{PhysRevX.8.031074} that quantifies the relationship between in-plane ($u, v$) and out-of-plane ($w$) components of the displacement field. The polarization factor is employed to discriminate and isolate out-of-plane polarized wave modes, which are weakly coupled to the in-plane ones. We compare the results to experimental data consisting of the out-of-plane velocity field $\dot{w}(x,y,t)$ of the plate's surface measured by a scanning laser Doppler vibrometer (SLDV) (see details in SM~\cite{SM}).

The numerically computed dispersion relations are shown in Fig.~\ref{Fig3a}, where, as previously, the red surface corresponds to the edge state, while the shaded gray volumes denote the bulk bands. Similar to the case of harmonic modulation, the existence of an edge state spanning the gap is associated with the non-trivial band topology identified by integer-valued Chern numbers. Chern numbers and gap labels are numerically evaluated using the Bloch modes obtained through the FE model according to the procedure outlined in SM~\cite{SM}, which yields the labels displayed in Fig.~\ref{Fig3a}. We find that the first spectral gap produced by the square modulation is also characterized by a gap label $C_g=1$, which signals a left to right transition of the edge state, again highlighted by the dashed and solid red lines at $\kappa_y=0$. This transition occurs at $\phi=\pi$ along the entire surface of the edge state, which is consistent with constant valued Chern numbers evaluated as a function of $\kappa_y$ in the considered frequency range ($f \in [0, 40]\;{\rm kHz}$). Figure~\ref{Fig3b} displays a zoomed view of the dispersion in Fig.~\ref{Fig3a} for $\phi \in [-0.7\pi,0.7\pi]$, with the observation that the interval $[-0.7\pi, 0]$ coincides with $[1.3\pi,2\pi]$ due to the periodicity with $\phi$. This interval corresponds to the phase modulation  of the manufactured plate, \textit{i.e.} from $\phi_i=0.7\pi$ to $\phi_f=-0.7\pi$, which exhibits the transition of the edge state occurring at the bottom boundary of the gap.

\begin{figure*}[ht!]
	\centering
	\subfigure[]{\includegraphics[height=0.25\textwidth]{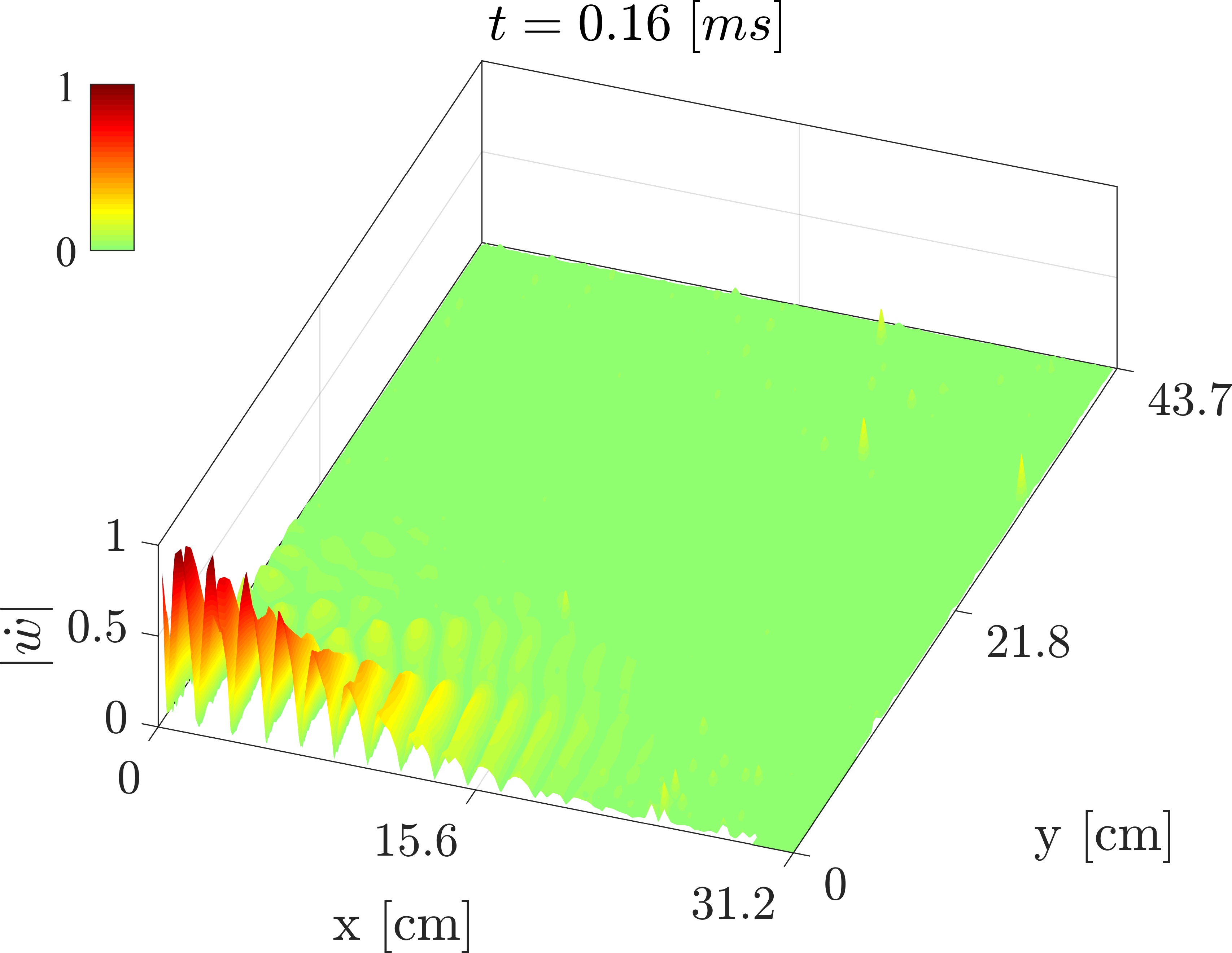}\label{Fig4a}}
	\subfigure[]{\includegraphics[height=0.25\textwidth]{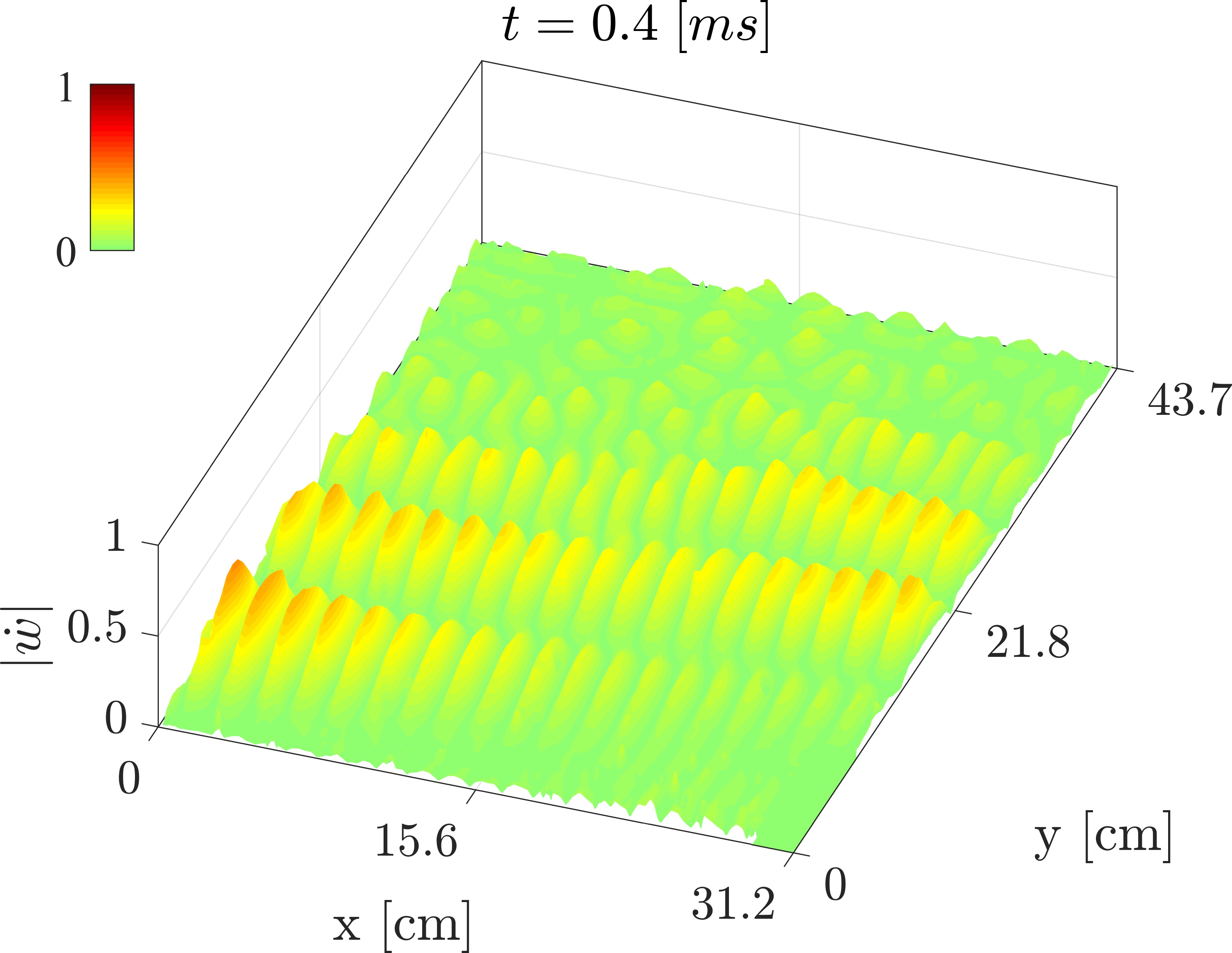}\label{Fig4b}}
	\subfigure[]{\includegraphics[height=0.25\textwidth]{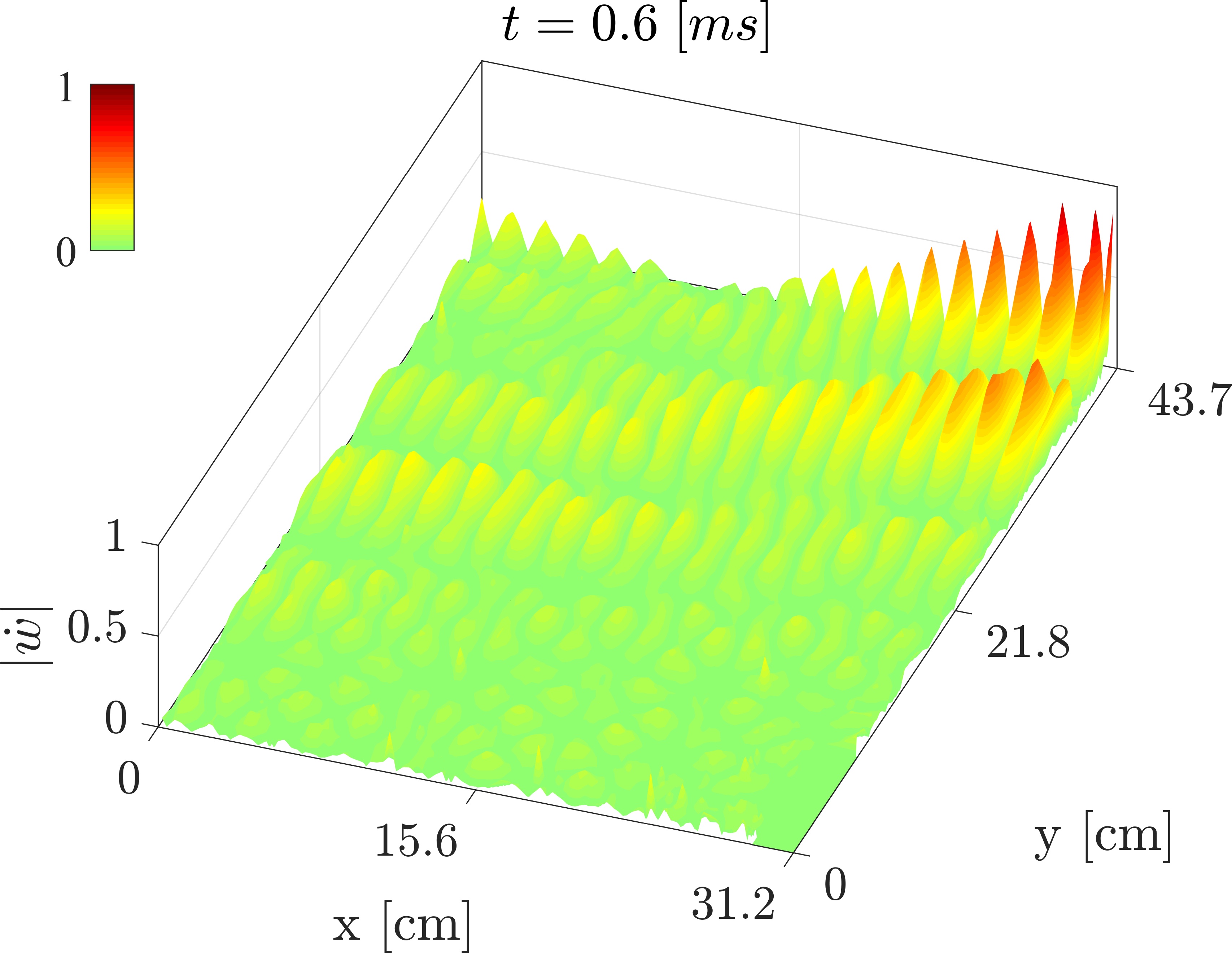}\label{Fig4c}}
	\caption{Experimental observation of transient topological pumping in square-modulated plate. Figures (a-c) displays snapshots of the measured velocity field for three subsequent time instants, where the transition from left-localized (a), to bulk (b), and finally to right localized mode (c) as the wave propagates along $y$ can be observed.}
	\label{Fig4}
\end{figure*}

The experiment is conducted by clamping the plate at its bottom right boundary while excitation is induced by a pair of piezoelectric ceramic patches attached to the bottom left boundary~\cite{SM}. The patches are connected to opposite electrical poles, which induces an out-of-phase (dipole) excitation that favors the excitation of the left-localized topological mode (for $\phi_i=0.7\pi$), while reducing the contribution from bulk modes. The forced frequency response function, corresponding to the response spatially averaged over the plate surface, is displayed alongside the dispersion surfaces. The finite size of the plate introduces a series of resonant peaks that are observed in the frequency range within which the edge state exists. At these frequencies, topological pumping is observed through a transition along the wavenumber branch of the edge state, as illustrated by the dashed and solid red lines in Fig.~\ref{Fig3b}, respectively denoting left and right localized modes. The measured velocity fields for three selected resonant frequencies are displayed in Figs.~\ref{Fig3}(c-e), where transitions from left to right localization characterizing the pump can be observed. For each of those recorded responses, a spectrogram is computed as described in Section~\ref{theorysec} and displayed in Figs.\ref{Fig3}(f-h), which confirms the transition along the wavenumber branch corresponding to the edge state. We remark that for the first peaks, such as in the case reported in Fig.\ref{Fig3}(c), the edge state is only defined within a restricted domain of the parameter space. In such cases, pumping still occurs through the wavenumber branch of the edge state, but is defined in a shorter spatial domain centered at the mid-portion of the plate, as confirmed by the spectrogram of Fig.\ref{Fig3}(f) corresponding to the pump of peak $II$.

The results reported in Fig.~\ref{Fig3} confirm the existence of steady-state topological pumping in the square-modulated plate occurring for several operating frequencies within $[25, 30]\;{\rm kHz}$, while populating the dispersion surface associated with the edge state. The potential of the modulated plate as a waveguiding platform is further demonstrated by realizing topological pumping in a transient regime. To that end, an excitation in the form of a $7-$cycle sine burst signal of center frequency $f=26.7$kHz is employed, which aims at transporting energy through the pump defined by peak $IV$ in Fig.~\ref{Fig3b}. Figure~\ref{Fig4} displays the measured velocity wavefield in the modulated plate at three subsequent time instants: a clear transition from left-localized wave (a), to bulk wave (b), and finally to right-localized wave propagation (c) is observed consistent with the expected topological pumping behavior. A video animation of the full transient response is provided in SM~\cite{SM}.

\section{Conclusions}\label{Conclusionsec}
In this paper, we present the first experimental demonstration of topological pumping in continuous elastic plates. We illustrate a simple design principle based on continuous property modulations which can be employed to induce the existence of topological edge states and drive their edge-to-edge transition. The results also provide opportunities for exploring higher dimensional physics in mechanics by exploiting synthetic dimensions in parameter space, which can be mapped to real spatial or temporal dimensions. These concepts have implications of technological relevance for applications involving elastic wave manipulation, such as guiding of bulk, surface and guided waves in acoustic devices, ultrasonic imaging and nondestructive evaluation.

\begin{acknowledgments}
The authors gratefully acknowledge the support from the National Science Foundation (NSF) through the EFRI 1741685 grant and from the Army Research office through grant W911NF-18-1-0036.
\end{acknowledgments}

\bibliographystyle{unsrt}
\bibliography{References}

\begin{thebibliography}{10}

\bibitem{hasan2010colloquium}
M~Zahid Hasan and Charles~L Kane.
\newblock Colloquium: topological insulators.
\newblock {\em Reviews of Modern Physics}, 82(4):3045, 2010.

\bibitem{lu2014topological}
Ling Lu, John~D Joannopoulos, and Marin Solja{\v{c}}i{\'c}.
\newblock Topological photonics.
\newblock {\em Nature Photonics}, 8(11):821, 2014.

\bibitem{khanikaev2013photonic}
Alexander~B Khanikaev, S~Hossein Mousavi, Wang-Kong Tse, Mehdi Kargarian,
  Allan~H MacDonald, and Gennady Shvets.
\newblock Photonic topological insulators.
\newblock {\em Nature materials}, 12(3):233, 2013.

\bibitem{PhysRevLett.114.114301}
Zhaoju Yang, Fei Gao, Xihang Shi, Xiao Lin, Zhen Gao, Yidong Chong, and Baile
  Zhang.
\newblock Topological acoustics.
\newblock {\em Phys. Rev. Lett.}, 114:114301, Mar 2015.

\bibitem{fleury2016floquet}
Romain Fleury, Alexander~B Khanikaev, and Andrea Alu.
\newblock Floquet topological insulators for sound.
\newblock {\em Nature communications}, 7:11744, 2016.

\bibitem{lu2017observation}
Jiuyang Lu, Chunyin Qiu, Liping Ye, Xiying Fan, Manzhu Ke, Fan Zhang, and
  Zhengyou Liu.
\newblock Observation of topological valley transport of sound in sonic
  crystals.
\newblock {\em Nature Physics}, 13(4):369, 2017.

\bibitem{mousavi2015topologically}
S~Hossein Mousavi, Alexander~B Khanikaev, and Zheng Wang.
\newblock Topologically protected elastic waves in phononic metamaterials.
\newblock {\em Nature communications}, 6:8682, 2015.

\bibitem{klitzing1980new}
K~v Klitzing, Gerhard Dorda, and Michael Pepper.
\newblock New method for high-accuracy determination of the fine-structure
  constant based on quantized hall resistance.
\newblock {\em Physical Review Letters}, 45(6):494, 1980.

\bibitem{thouless1982quantized}
David~J Thouless, Mahito Kohmoto, M~Peter Nightingale, and Md~den Nijs.
\newblock Quantized hall conductance in a two-dimensional periodic potential.
\newblock {\em Physical review letters}, 49(6):405, 1982.

\bibitem{prodan2009topological}
Emil Prodan and Camelia Prodan.
\newblock Topological phonon modes and their role in dynamic instability of
  microtubules.
\newblock {\em Physical review letters}, 103(24):248101, 2009.

\bibitem{wang2015topological}
Pai Wang, Ling Lu, and Katia Bertoldi.
\newblock Topological phononic crystals with one-way elastic edge waves.
\newblock {\em Physical review letters}, 115(10):104302, 2015.

\bibitem{nash2015topological}
Lisa~M Nash, Dustin Kleckner, Alismari Read, Vincenzo Vitelli, Ari~M Turner,
  and William~TM Irvine.
\newblock Topological mechanics of gyroscopic metamaterials.
\newblock {\em Proceedings of the National Academy of Sciences},
  112(47):14495--14500, 2015.

\bibitem{souslov2017topological}
Anton Souslov, Benjamin~C Van~Zuiden, Denis Bartolo, and Vincenzo Vitelli.
\newblock Topological sound in active-liquid metamaterials.
\newblock {\em Nature Physics}, 13(11):1091, 2017.

\bibitem{mitchell2018amorphous}
Noah~P Mitchell, Lisa~M Nash, Daniel Hexner, Ari~M Turner, and William~TM
  Irvine.
\newblock Amorphous topological insulators constructed from random point sets.
\newblock {\em Nature Physics}, 14(4):380, 2018.

\bibitem{chen2019mechanical}
H~Chen, LY~Yao, H~Nassar, and GL~Huang.
\newblock Mechanical quantum hall effect in time-modulated elastic materials.
\newblock {\em Physical Review Applied}, 11(4):044029, 2019.

\bibitem{susstrunk2015observation}
Roman S{\"u}sstrunk and Sebastian~D Huber.
\newblock Observation of phononic helical edge states in a mechanical
  topological insulator.
\newblock {\em Science}, 349(6243):47--50, 2015.

\bibitem{pal2016helical}
Raj~Kumar Pal, Marshall Schaeffer, and Massimo Ruzzene.
\newblock Helical edge states and topological phase transitions in phononic
  systems using bi-layered lattices.
\newblock {\em Journal of Applied Physics}, 119(8):084305, 2016.

\bibitem{PhysRevB.98.094302}
H.~Chen, H.~Nassar, A.~N. Norris, G.~K. Hu, and G.~L. Huang.
\newblock Elastic quantum spin hall effect in kagome lattices.
\newblock {\em Phys. Rev. B}, 98:094302, Sep 2018.

\bibitem{chaunsali2018subwavelength}
Rajesh Chaunsali, Chun-Wei Chen, and Jinkyu Yang.
\newblock Subwavelength and directional control of flexural waves in
  zone-folding induced topological plates.
\newblock {\em Physical Review B}, 97(5):054307, 2018.

\bibitem{PhysRevX.8.031074}
M.~Miniaci, R.~K. Pal, B.~Morvan, and M.~Ruzzene.
\newblock Experimental observation of topologically protected helical edge
  modes in patterned elastic plates.
\newblock {\em Phys. Rev. X}, 8:031074, Sep 2018.

\bibitem{pal2017edge}
Raj~Kumar Pal and Massimo Ruzzene.
\newblock Edge waves in plates with resonators: an elastic analogue of the
  quantum valley hall effect.
\newblock {\em New Journal of Physics}, 19(2):025001, 2017.

\bibitem{vila2017observation}
Javier Vila, Raj~Kumar Pal, and Massimo Ruzzene.
\newblock Observation of topological valley modes in an elastic hexagonal
  lattice.
\newblock {\em Physical Review B}, 96(13):134307, 2017.

\bibitem{liu2018tunable}
Ting-Wei Liu and Fabio Semperlotti.
\newblock Tunable acoustic valley--hall edge states in reconfigurable phononic
  elastic waveguides.
\newblock {\em Physical Review Applied}, 9(1):014001, 2018.

\bibitem{liu2019experimental}
Ting-Wei Liu and Fabio Semperlotti.
\newblock Experimental evidence of robust acoustic valley hall edge states in a
  nonresonant topological elastic waveguide.
\newblock {\em Physical Review Applied}, 11(1):014040, 2019.

\bibitem{qi2008topological}
Xiao-Liang Qi, Taylor~L Hughes, and Shou-Cheng Zhang.
\newblock Topological field theory of time-reversal invariant insulators.
\newblock {\em Physical Review B}, 78(19):195424, 2008.

\bibitem{kraus2016quasiperiodicity}
Yaacov~E Kraus and Oded Zilberberg.
\newblock Quasiperiodicity and topology transcend dimensions.
\newblock {\em Nature Physics}, 12(7):624, 2016.

\bibitem{ozawa2016synthetic}
Tomoki Ozawa, Hannah~M Price, Nathan Goldman, Oded Zilberberg, and Iacopo
  Carusotto.
\newblock Synthetic dimensions in integrated photonics: From optical isolation
  to four-dimensional quantum hall physics.
\newblock {\em Physical Review A}, 93(4):043827, 2016.

\bibitem{lee2018electromagnetic}
Ching~Hua Lee, Yuzhu Wang, Youjian Chen, and Xiao Zhang.
\newblock Electromagnetic response of quantum hall systems in dimensions five
  and six and beyond.
\newblock {\em Physical Review B}, 98(9):094434, 2018.

\bibitem{alvarez2019edge}
VM~Martinez Alvarez and MD~Coutinho-Filho.
\newblock Edge states in trimer lattices.
\newblock {\em Physical Review A}, 99(1):013833, 2019.

\bibitem{kraus2012topological}
Yaacov~E Kraus, Yoav Lahini, Zohar Ringel, Mor Verbin, and Oded Zilberberg.
\newblock Topological states and adiabatic pumping in quasicrystals.
\newblock {\em Physical review letters}, 109(10):106402, 2012.

\bibitem{apigo2019observation}
David~J Apigo, Wenting Cheng, Kyle~F Dobiszewski, Emil Prodan, and Camelia
  Prodan.
\newblock Observation of topological edge modes in a quasiperiodic acoustic
  waveguide.
\newblock {\em Physical review letters}, 122(9):095501, 2019.

\bibitem{ni2019observation}
Xiang Ni, Kai Chen, Matthew Weiner, David~J Apigo, Camelia Prodan, Andrea
  Al{\`u}, Emil Prodan, and Alexander~B Khanikaev.
\newblock Observation of hofstadter butterfly and topological edge states in
  reconfigurable quasi-periodic acoustic crystals.
\newblock {\em Communications Physics}, 2(1):55, 2019.

\bibitem{apigo2018topological}
David~J Apigo, Kai Qian, Camelia Prodan, and Emil Prodan.
\newblock Topological edge modes by smart patterning.
\newblock {\em Physical Review Materials}, 2(12):124203, 2018.

\bibitem{rosa2019edge}
Matheus~IN Rosa, Raj~Kumar Pal, Jos{\'e}~RF Arruda, and Massimo Ruzzene.
\newblock Edge states and topological pumping in spatially modulated elastic
  lattices.
\newblock {\em Physical Review Letters}, 123(3):034301, 2019.

\bibitem{Pal_2019}
Raj~Kumar Pal, Matheus I~N Rosa, and Massimo Ruzzene.
\newblock Topological bands and localized vibration modes in quasiperiodic
  beams.
\newblock {\em New Journal of Physics}, 21(9):093017, sep 2019.

\bibitem{harper1955single}
Philip~George Harper.
\newblock Single band motion of conduction electrons in a uniform magnetic
  field.
\newblock {\em Proceedings of the Physical Society. Section A}, 68(10):874,
  1955.

\bibitem{aubry1980analyticity}
Serge Aubry and Gilles Andr{\'e}.
\newblock Analyticity breaking and anderson localization in incommensurate
  lattices.
\newblock {\em Ann. Israel Phys. Soc}, 3(133):18, 1980.

\bibitem{zilberberg2018photonic}
Oded Zilberberg, Sheng Huang, Jonathan Guglielmon, Mohan Wang, Kevin~P Chen,
  Yaacov~E Kraus, and Mikael~C Rechtsman.
\newblock Photonic topological boundary pumping as a probe of 4d quantum hall
  physics.
\newblock {\em Nature}, 553(7686):59, 2018.

\bibitem{lohse2018exploring}
Michael Lohse, Christian Schweizer, Hannah~M Price, Oded Zilberberg, and
  Immanuel Bloch.
\newblock Exploring 4d quantum hall physics with a 2d topological charge pump.
\newblock {\em Nature}, 553(7686):55, 2018.

\bibitem{petrides2018six}
Ioannis Petrides, Hannah~M Price, and Oded Zilberberg.
\newblock Six-dimensional quantum hall effect and three-dimensional topological
  pumps.
\newblock {\em Physical Review B}, 98(12):125431, 2018.

\bibitem{verbin2015topological}
Mor Verbin, Oded Zilberberg, Yoav Lahini, Yaacov~E Kraus, and Yaron Silberberg.
\newblock Topological pumping over a photonic fibonacci quasicrystal.
\newblock {\em Physical Review B}, 91(6):064201, 2015.

\bibitem{nakajima2016topological}
Shuta Nakajima, Takafumi Tomita, Shintaro Taie, Tomohiro Ichinose, Hideki
  Ozawa, Lei Wang, Matthias Troyer, and Yoshiro Takahashi.
\newblock Topological thouless pumping of ultracold fermions.
\newblock {\em Nature Physics}, 12(4):296, 2016.

\bibitem{lohse2016thouless}
Michael Lohse, Christian Schweizer, Oded Zilberberg, Monika Aidelsburger, and
  Immanuel Bloch.
\newblock A thouless quantum pump with ultracold bosonic atoms in an optical
  superlattice.
\newblock {\em Nature Physics}, 12(4):350, 2016.

\bibitem{grinberg2019robust}
Inbar~Hotzen Grinberg, Mao Lin, Cameron Harris, Wladimir~A Benalcazar,
  Christopher~W Peterson, Taylor~L Hughes, and Gaurav Bahl.
\newblock Robust temporal pumping in a magneto-mechanical topological
  insulator.
\newblock {\em arXiv preprint arXiv:1905.02778}, 2019.

\bibitem{graff2012wave}
Karl~F Graff.
\newblock {\em Wave motion in elastic solids}.
\newblock Courier Corporation, 2012.

\bibitem{SM}
See supplemental material at xxxx for more details on the simulation
  procedures, experimental setup and methodology, and for a video animation of
  the transient experimental analysis.

\bibitem{meirovitch1975elements}
L~Meirovitch.
\newblock {\em Elements of vibration analysis}.
\newblock McGraw-Hill, 1975.

\bibitem{hatsugai1993chern}
Yasuhiro Hatsugai.
\newblock Chern number and edge states in the integer quantum hall effect.
\newblock {\em Physical review letters}, 71(22):3697, 1993.

\bibitem{fukui2005chern}
Takahiro Fukui, Yasuhiro Hatsugai, and Hiroshi Suzuki.
\newblock Chern numbers in discretized brillouin zone: efficient method of
  computing (spin) hall conductances.
\newblock {\em Journal of the Physical Society of Japan}, 74(6):1674--1677,
  2005.

\bibitem{nassar2018quantization}
H~Nassar, H~Chen, AN~Norris, and GL~Huang.
\newblock Quantization of band tilting in modulated phononic crystals.
\newblock {\em Physical Review B}, 97(1):014305, 2018.

\end{thebibliography}
\end{document}